\documentclass[pdflatex,sn-mathphys-num,iicol]{sn-jnl}

\usepackage{graphicx}
\usepackage{amsmath,amssymb,amsfonts}
\usepackage{mathtools}
\usepackage{bm}
\usepackage{booktabs}
\usepackage{xcolor}
\usepackage{textcomp}

\begin{document}

\title[Finite-field QED in magnetars]{Finite-Field QED Corrections to Vacuum Birefringence and Magnetar Polarization Transport\textsuperscript{\(\dagger\)}}

\author*[1]{\fnm{Shahram} \sur{Abbassi}}\email{sabbassi@uwo.ca}
\author[2,3]{\fnm{F. A.} \sur{Chishtie}}
\author[1]{\fnm{S. R.} \sur{Valluri}}

\affil*[1]{\orgdiv{Department of Physics and Astronomy}, \orgname{The University of Western Ontario}, \orgaddress{\city{London}, \state{Ontario}, \postcode{N6A 3K7}, \country{Canada}}}
\affil[2]{\orgname{Peaceful Society, Science and Innovation Foundation}, \orgaddress{\city{Vancouver}, \state{British Columbia}, \postcode{V6K 2E4}, \country{Canada}}}
\affil[3]{\orgdiv{Department of Occupational Science and Occupational Therapy}, \orgname{The University of British Columbia}, \orgaddress{\city{Vancouver}, \state{British Columbia}, \postcode{V6T 2L4}, \country{Canada}}}

\abstract{We study low-energy photon propagation in a constant magnetic field within the one-loop Heisenberg--Euler theory, retaining the refractive-index normalization $\gamma_s$ without expansion. Here ``finite-field'' denotes exact dependence on $B/B_{\rm cr}$ within the one-loop, constant-field approximation. The resulting birefringence is propagated into magnetar polarization transport. In a centered-dipole model, the polarization-limiting radius is unchanged to better than $10^{-12}$ because mode decoupling occurs at $\sim10^2R_{\rm NS}$, where $B\ll B_{\rm cr}$. Near the surface, however, the weak-field Cotton--Mouton expression overestimates the accumulated birefringent phase by up to a factor $2.9$ at $10^{15}$~G. At the plasma--vacuum resonance, finite-field corrections reduce the resonance density by $32\%$ and raise the adiabatic conversion energy by $14\%$ for 1E~1547.0$-$5408; the corresponding changes are factors $2.6$ and $1.37$ for 1RXS~J1708$-$4009, and factors $9.7$ and $2.13$ for SGR~1806$-$20, the latter controlled by the strong-field asymptote. The resummed one-loop parallel-mode magnetic response remains positive and develops a broad maximum near $17B_{\rm cr}$. The strictly truncated $\mathcal O(\alpha)$ response is monotonic; therefore the maximum is a structural prediction of the resummed one-loop constitutive model, while its detailed profile and precise location require higher-loop validation. These results identify vacuum-resonance observables as the most sensitive channel for testing finite-field QED in magnetars.}


\maketitle
\begingroup
\renewcommand{\thefootnote}{\fnsymbol{footnote}}
\footnotetext[2]{The authors gratefully acknowledge the late Dr. J. W. Mielniczuk, who passed away in July 2016, for his valuable contributions to the analytical development and early numerical verification of this work.}

\endgroup

%===========================================================================
\section{Introduction} 
\label{sec:intro}
%===========================================================================

The nonlinearity of Maxwell's equations in the quantum regime continues to present us with a variety of fascinating phenomena. Born and Infeld (1934) pioneered nonlinear electrodynamics from the perspective of classical field theory \cite{bif34,dun12}. In quantum electrodynamics (QED), nonlinear electromagnetic effects arise naturally from the polarization of the vacuum by virtual charged particles, in particular electron--positron pairs \cite{bai0,dit00}. These effects imply that the quantum vacuum behaves as a nonlinear optical medium when probed by sufficiently strong electromagnetic fields. More recently, several aspects of this magnetized quantum vacuum have been studied by \cite{sha11}, \cite{cha12}, and \cite{alt08}, among others.

In the low-energy regime, where the characteristic photon frequency is well below the pair-creation threshold and the external electromagnetic field varies slowly on the scale of the electron Compton wavelength, the one-loop Heisenberg--Euler Lagrangian (HEL) provides the standard effective-field-theory description of the QED vacuum \cite{hei36,mk79,sch51,dun04,sha11,cha12}. The HEL predicts several closely related nonlinear phenomena, including vacuum birefringence,photon--photon scattering, vacuum polarization, and field-dependent modifications of photon propagation. These effects are very small in terrestrial magnetic fields, but they become increasingly important in ultra-strong fields comparable to the Schwinger critical field $B_{\rm cr}=m^2/e$, where the magnetic-field energy scale is set by the electron mass.

Vacuum birefringence is one of the cleanest manifestations of the nonlinear QED vacuum. In a magnetized vacuum, the two physical photon eigenmodes acquire different refractive indices, so that the propagation of light depends on its polarization and on the angle between the photon wave vector and the external magnetic field. This anisotropic response makes the vacuum analogous to a birefringent optical medium, but with the anisotropy generated entirely by quantum fluctuations. Theoretical studies of this effect have a long history, ranging from the original Heisenberg--Euler theory to photon propagation and photon splitting in external fields \cite{bia70} and later analyses of photon propagation in strong magnetic fields \cite{hei36,bai0,dit00,hey97a,hey97b,hey97c,sha11,cha12}. Early examples of this nonlinear optical structure include the demonstration that higher harmonics arise from the HEL in the ultra-strong \cite{bha78} and weak \cite{val80} field limits, and the study of the paramagnetic properties of the strongly magnetized QED vacuum \cite{mie88}.

The experimental search for vacuum nonlinearities has advanced significantly in recent years. Laboratory polarimetry experiments such as PVLAS aim to detect the tiny ellipticity induced by magnetic vacuum birefringence\cite{zav08,bre08,can08,val15,pvlas20}. After decades of development, PVLAS has reached sensitivities close to the QED prediction, with current measurements giving $\Delta n=(12\pm17)\times10^{-23}$ at 
$B=2.5$ T \cite{pvlas20}. Future laboratory searches are also being developed at high-intensity laser and X-ray facilities, including the BIREF@HIBEF program at the European XFEL, which aims to probe vacuum birefringence using an X-ray probe beam in a strong optical-laser background \cite{birefhibef25}. Collider measurements provide a complementary probe of the same nonlinear QED framework: the ATLAS observation of light-by-light scattering in ultra-peripheral heavy-ion collisions, with a significance of $8.2\sigma$, confirms photon--photon scattering in a high-energy regime \cite{atlas17,atlas19}. Although this collider process is not a direct measurement of the static-field photon magnetic moment considered here, it provides important experimental support for the nonlinear QED vacuum response.

Astrophysical systems provide another route to the strong-field regime. Magnetars, whose surface magnetic fields can reach $10^{14}$--$10^{15}$ Gauss, are natural laboratories for QED vacuum birefringence \cite{isp02}. Optical polarimetry of the isolated neutron star RX J1856.5-3754 has provided suggestive evidence for vacuum birefringence effects \cite{mig17}, and related astrophysical implications were examined in Ref.~\cite{val17}. More recently, X-ray polarimetry with IXPE has opened a new observational window on magnetized neutron-star magnetospheres \cite{taverna22,lai23,ixpe25}. These observations are highly relevant to the physics of polarization transport in ultra-strong magnetic fields, although their interpretation requires detailed modeling of emission geometry, plasma effects, magnetospheric structure, and radiative transfer.

While vacuum birefringence has been studied extensively, the photon anomalous magnetic moment provides a complementary way of characterizing the same field-dependent dispersion physics. Unlike the electron anomalous magnetic moment, which arises from radiative corrections to a charged particle, the photon magnetic moment in a magnetized vacuum is an effective property of photon propagation induced by vacuum polarization. Its paramagnetic behavior and limiting forms have been studied by P\'erez Rojas \& Rodr\'iguez Querts \cite{roj06,roj07,roj14}, and related aspects of photon propagation and angular momentum in magnetized vacuum have been discussed by Villalba-Ch\'avez \& Shabad \cite{cha12}. These works establish the physical basis for assigning a magnetic response to photon eigenmodes in an external magnetic field.

The remaining gap is that most analytic discussions of the photon magnetic moment emphasize limiting regimes -- weak fields or asymptotically strong fields -- whereas the intermediate domain $B\sim B_{\rm cr}$, where neither expansion alone suffices, is precisely the regime that connects laboratory birefringence, magnetar-scale applications, and the weak-to-strong-field transition. In this paper we derive and numerically validate finite-field expressions for the photon magnetic response over $0\leq B\leq30\,B_{\rm cr}$ within the one-loop low-energy framework, keeping the refractive-index normalization $\gamma_s$ exact and resolving the polarization and angular dependence, so that the commonly used 
perpendicular-propagation result appears as a special case of the general anisotropic dispersion. The goal is not to claim a qualitatively new nonlinear-QED effect but to provide an internally consistent finite-field description of a known 
one, suitable as quantitative input for polarization observables.

This finite-field treatment has consequences that go beyond the underlying one-loop formulas. Keeping $\gamma_s$ exact reveals that the normalized parallel-mode response $\widehat{\mu}_{\gamma}^{(\parallel)}(\xi)$ is not globally 
monotonic over $0\leq\xi\leq30$: it reaches a broad maximum near $\xi\simeq17$, which Sec.~\ref{sec:peak_analytic} and Appendix~\ref{app:positivity} characterize analytically through a general extremum condition, located exactly and reproduced to $1.4\%$ by a closed-form estimate built from the strong-field expansion of Sec.~\ref{sec:propagation}. Section~\ref{sec:peak_analytic} also assesses the perturbative status of this feature: the strictly truncated 
$\mathcal{O}(\alpha)$ response is monotonic, and the maximum is generated by the resummed normalization at the $\sim2\%$ level, comparable in formal order to omitted two-loop Heisenberg--Euler corrections \cite{rit75,gie17,kar19}, so its existence and approximate location -- not its precise profile -- are the controlled statements. The corresponding field scale, $B\simeq7.5\times10^{14}$~G, lies within the observed range of magnetar surface fields, so the feature sits in an astrophysically realized regime rather than a formal one. On the observable side, the same exact $\Delta n$ delimits precisely where finite-field input is required in polarization-transport modeling. The polarization-limiting radius $r_{\rm pl}$ of magnetars proves insensitive to the correction -- mode decoupling occurs far outside the supercritical region -- while the weak-field Cotton--Mouton formula overestimates the birefringent phase accumulated near the surface by up to a factor $\sim3$ for $B_{s}=10^{15}$~G (Sec.~\ref{sec:rpl}). At the plasma--vacuum resonance, by contrast, all atmosphere-dependent factors cancel in the exact-to-weak-field ratios, and resonance-based observables become sensitive to the finite-field correction for surface fields $\xi_{s}\gtrsim10$ -- including 1RXS~J1708$-$4009, the leading near-term polarimetric target (Sec.~\ref{sec:vres}).

The paper is organized as follows. In Section~\ref{sec:theory}, we establish the one-loop Heisenberg--Euler framework, define the polarization modes and propagation geometry, and fix the refractive-index normalization convention. In 
Section~\ref{sec:propagation}, we derive the polarization-dependent refractive indices, discuss the weak- and strong-field limits, and connect the birefringent index splitting to ellipticity. In Section~\ref{sec:mu}, we formulate the 
finite-field photon magnetic response in terms of the field dependence of the polarization-dependent dispersion relation, derive its weak- and strong-field limits, and clarify its validity and physical interpretation. In Section~\ref{sec:numerics_context}, we present the numerical validation of the finite-field response, summarize the laboratory and astrophysical context, and quantify the finite-field corrections to the magnetar polarization-limiting radius and to the plasma--vacuum resonance (Secs.~\ref{sec:rpl} and \ref{sec:vres}). Section~\ref{sec:conclusions} contains our conclusions. Supplementary mathematical details are provided in Appendices A, B, and C.

%===========================================================================
\section{Heisenberg--Euler Framework}
\label{sec:theory}
%===========================================================================

\subsection{Effective Lagrangian}

At one-loop order, the Heisenberg--Euler effective Lagrangian in constant external electromagnetic fields \cite{hei36,kar15} describes the nonlinear response of the QED vacuum induced by virtual electron--positron fluctuations. In the proper-time representation of Schwinger \cite{sch51}, the one-loop contribution can be written as
\begin{equation}
\label{eq:Lagrangian}
\begin{split}
\mathcal{L}^{(1)}
&= \frac{\alpha}{2\pi}\int_0^{\infty}\frac{ds}{s}\,
\exp\!\left(-i\frac{m^2}{e}s\right)  \\
&\quad \times
\left[
ab\,\coth(as)\cot(bs)
-\frac{a^2-b^2}{3}
-\frac{1}{s^2}
\right] .
\end{split}
\end{equation}
with the prescription $m^2 \rightarrow m^2-i0^{+}$, and with the proper-time integration contour taken slightly below the real positive $s$ axis. Here, $m$ is the electron mass, $e$ is the elementary charge, and $\alpha=e^2/(4\pi)$ is the fine-structure constant. The full effective Lagrangian used in the following derivations is
\begin{equation}
\label{eq:Leff}
{
\mathcal{L}_{\rm eff}
=
-\mathcal{F}
+
\mathcal{L}^{(1)} ,
}
\end{equation}
where the first term is the classical Maxwell contribution. This distinction is important because derivatives such as $\partial\mathcal{L}_{\rm eff}/\partial\mathcal{F}$ contain the leading Maxwell term as well as the one-loop QED correction. In what follows, $\mathcal{L}$ denotes $\mathcal{L}_{\rm eff}$ unless otherwise stated.

The quantities $a$ and $b$ are the secular invariants constructed from the gauge- and Lorentz-invariant combinations of the electromagnetic field,
\begin{subequations}
\label{eq:secular_invariants}
\begin{align}
a &=
\left[
\left(\mathcal{F}^2+\mathcal{G}^2\right)^{1/2}
-\mathcal{F}
\right]^{1/2}, \\
b &=
\left[
\left(\mathcal{F}^2+\mathcal{G}^2\right)^{1/2}
+\mathcal{F}
\right]^{1/2},
\end{align}
\end{subequations}
where
\begin{subequations}
\label{eq:field_invariants}
\begin{align}
\mathcal{F}
&=
\frac{1}{4}F_{\mu\nu}F^{\mu\nu}
=
\frac{1}{2}
\left(\mathbf{B}^2-\mathbf{E}^2\right), \\
\mathcal{G}
&=
\frac{1}{4}F^*_{\mu\nu}F^{\mu\nu}
=
-\mathbf{E}\cdot\mathbf{B}.
\end{align}
\end{subequations}
Here $F^{*\mu\nu}= \frac{1}{2}\epsilon^{\mu\nu\alpha\beta}F_{\alpha\beta}$ is the dual field-strength tensor, with $\epsilon^{0123}=1$. We use the metric convention $g_{\mu\nu}=\mathrm{diag}(-1,+1,+1,+1)$ and natural units 
$c=\hbar=1$.

\subsection{Polarization Modes and Geometry}

We consider a photon with wave vector $\mathbf{k}$ propagating through a region with constant external magnetic field $\mathbf{B}$ and zero external electric field, 
$\mathbf{E}=0$. Thus,
\begin{equation}
\mathcal{G}=0, 
\qquad 
\mathcal{F}=\frac{B^2}{2},
\end{equation}
where $B\equiv|\mathbf{B}|$ denotes the magnitude of the external magnetic field. The angle $\theta$ is defined by
\begin{equation}
\cos\theta
=
\frac{\mathbf{B}\cdot\mathbf{k}}{|\mathbf{B}|\,|\mathbf{k}|}.
\label{eq:theta_def}
\end{equation}

The two physical photon eigenmodes are defined with respect to the plane spanned by $\mathbf{B}$ and $\mathbf{k}$. We use the labels $\perp$ and $\parallel$ to refer to the orientation of the photon's electric field $\mathbf{E}_\gamma$ relative to this plane:
\begin{itemize}
\item \textbf{Perpendicular mode ($\perp$):} $\mathbf{E}_\gamma$ is perpendicular to 
the $(\mathbf{B},\mathbf{k})$ plane. Equivalently, the photon magnetic field 
$\mathbf{B}_\gamma$ lies in this plane.
\item \textbf{Parallel mode ($\parallel$):} $\mathbf{E}_\gamma$ lies in the 
$(\mathbf{B},\mathbf{k})$ plane.
\end{itemize}

This labeling follows the convention of Adler \cite{adl71} and of the neutron-star polarization-transport literature \cite{hey97a,lai02}: for $\theta=\pi/2$ the $\parallel$ mode has $\mathbf{E}_\gamma$ parallel to the external 
field and corresponds to the ordinary (O) mode, while the $\perp$ mode has $\mathbf{E}_\gamma$ perpendicular to the external field and corresponds to the extraordinary (X) mode. Because the $\parallel$ mode couples to the invariant 
$\mathcal{G}=-\mathbf{E}\cdot\mathbf{B}$ while the $\perp$ mode couples through $\mathcal{F}$, the $\parallel$ mode carries the larger weak-field coefficient [$14/45$ versus $8/45$ in Eqs.~(\ref{eq:n_perp_weak})--(\ref{eq:n_parallel_weak}) 
below]. We caution that the opposite labeling, in which the modes are classified by the orientation of the photon magnetic field $\mathbf{B}_\gamma$ rather than $\mathbf{E}_\gamma$, also appears in the strong-field literature; with that 
convention the two weak-field coefficients are interchanged. All labels in this paper refer to $\mathbf{E}_\gamma$.

These two modes acquire different refractive indices in a magnetized QED vacuum, leading to vacuum birefringence. The external magnetic field selects a preferred spatial direction and therefore breaks the isotropy of the vacuum, making photon propagation analogous to propagation in an anisotropic optical medium. For propagation exactly parallel to the external field, $\theta=0$, the $(\mathbf{B},\mathbf{k})$ plane is degenerate and the two physical polarizations are 
equivalent; correspondingly, the birefringent response vanishes in the standard one-loop magnetic-vacuum expressions through the factor $\sin^2\theta$.

\subsection{Field Derivatives}

For a purely magnetic external background, $\mathbf{E}=0$, the derivatives of the effective Lagrangian that enter the photon dispersion relations are evaluated at $\mathcal{F}=B^2/2$ and $\mathcal{G}=0$. Since the full effective Lagrangian 
was defined in Eq.~(\ref{eq:Leff}) as
\[
\mathcal{L}_{\rm eff}
=
-\mathcal{F}
+
\mathcal{L}^{(1)},
\]
it is important to distinguish the derivative of the full effective Lagrangian from the derivatives of the one-loop correction. We define
\begin{equation}
\label{eq:LF_full_def}
\mathcal{L}_{\mathcal{F}}
\equiv
\left.
\frac{\partial \mathcal{L}_{\rm eff}}{\partial \mathcal{F}}
\right|_{\mathcal{F}=B^2/2,\,\mathcal{G}=0}
\end{equation}
and reserve the notation $\gamma_{\mathcal{F}}$ for the one-loop correction,
\begin{equation}
\label{eq:gamma_defs}
\begin{split}
\gamma_{\mathcal{F}}
&\equiv 
\left.
\frac{\partial \mathcal{L}^{(1)}}{\partial \mathcal{F}}
\right|_{\mathcal{F}=B^2/2,\,\mathcal{G}=0},  \\
\gamma_{\mathcal{FF}}
&\equiv 
\left.
\frac{\partial^2 \mathcal{L}^{(1)}}{\partial \mathcal{F}^2}
\right|_{\mathcal{F}=B^2/2,\,\mathcal{G}=0},  \\
\gamma_{\mathcal{GG}}
&\equiv 
\left.
\frac{\partial^2 \mathcal{L}^{(1)}}{\partial \mathcal{G}^2}
\right|_{\mathcal{F}=B^2/2,\,\mathcal{G}=0}.
\end{split}
\end{equation}
The Maxwell term contributes only to the first derivative, so that
\begin{equation}
\label{eq:LF_gamma_relation}
\mathcal{L}_{\mathcal{F}}
=
-1+\gamma_{\mathcal{F}} .
\end{equation}
The second derivatives $\gamma_{\mathcal{FF}}$ and $\gamma_{\mathcal{GG}}$ are therefore purely one-loop quantities at the order considered here. This convention is essential for obtaining the standard weak-field refractive-index coefficients. Because the background is purely magnetic, terms that are odd in $\mathcal{G}$, such as $\gamma_{\mathcal{G}}$ and $\gamma_{\mathcal{FG}}$, vanish.

We introduce the dimensionless parameter
\begin{equation}
\label{eq:h_def}
h=\frac{1}{2}\frac{B_{\rm cr}}{B},
\end{equation}
where
\begin{equation}
B_{\rm cr}=\frac{m^2}{e}
\simeq 4.414\times10^{9}\,{\rm T}
\simeq 4.414\times10^{13}\,{\rm G}
\end{equation}
is the Schwinger critical field. The zero-field limit corresponds to $h\rightarrow\infty$ and is always understood in this limiting sense. In terms of $h$, the one-loop derivatives may be written as \cite{lun09,lun10}
\begin{align}
\gamma_{\mathcal{F}}
&=
-\frac{\alpha}{2\pi}
\bigg[
\frac{1}{3}
+2h^2
-8\zeta^{\prime}(-1,h)
+4h\ln\Gamma(h)
\nonumber\\
&\hspace{1.6cm}
-2h\ln h
+\frac{2}{3}\ln h
-2h\ln(2\pi)
\bigg],
\label{eq:gammaF}
\end{align}
\begin{align}
\gamma_{\mathcal{FF}}
&=
\frac{\alpha}{2\pi B^2}
\bigg[
\frac{2}{3}
+4h^2\psi(1+h)
-2h
\nonumber\\
&\hspace{1.25cm}
-4h^2
-4h\ln\Gamma(h)
\nonumber\\
&\hspace{1.25cm}
+2h\ln(2\pi)
-2h\ln h
\bigg],
\label{eq:gammaFF}
\end{align}
and
\begin{align}
\gamma_{\mathcal{GG}}
&=
\frac{\alpha}{2\pi B^2}
\bigg[
\frac{1}{3}
\left(
\frac{1}{h}-1
\right)
-\frac{2}{3}\psi(1+h)
\nonumber\\
&\hspace{1.25cm}
-2h^2
+8\zeta^{\prime}(-1,h)
\nonumber\\
&\hspace{1.25cm}
-4h\ln\Gamma(h)
+2h\ln(2\pi h)
\bigg].
\label{eq:gammaGG}
\end{align}
Here $\psi(x)=\Gamma'(x)/\Gamma(x)$ is the digamma function, $\Gamma(x)$ is the Euler gamma function, and
\begin{equation}
\zeta^{\prime}(s,h)\equiv \partial_s\zeta(s,h),
\end{equation}
with $\zeta(s,h)$ the Hurwitz zeta function.

For $s=-1$ and $h\gg1$, one may use the representation \cite{ada04,dit79}
\begin{align}
\zeta^{\prime}(-1,h)
&\simeq
\frac{1}{12}
-\frac{h^2}{4}
+\frac{\ln h}{2}
\left(
h^2-h+\frac{1}{6}
\right)
\nonumber\\
&\quad
+
\int^{\infty}_{0}
\frac{e^{-hx}}{x^2}
\bigg[
\frac{1}{1-e^{-x}}
-\frac{1}{x}
\nonumber\\
&\hspace{3.0cm}
-\frac{1}{2}
-\frac{x}{12}
\bigg]dx ,
\label{eq:zeta_integral}
\end{align}
valid for $\mathrm{Re}(h)>0$. The corresponding asymptotic approximation is
\begin{equation}
\label{eq:zeta_approx}
\zeta^{\prime}(-1,h)
\simeq
\frac{1}{12}
-\frac{h^2}{4}
+\frac{\ln h}{2}B_2(h)
+\frac{1}{720h^2},
\end{equation}
where $B_2(h)=h^2-h+1/6$ is the second Bernoulli polynomial \cite{olv10}. The integral 
in Eq.~(\ref{eq:zeta_integral}) is convergent \cite{ada04}.

We finally define the refractive-index normalization factor by
\begin{equation}
\label{eq:gamma_s_def}
\gamma_s
\equiv
-\mathcal{L}_{\mathcal{F}}
=
1-\gamma_{\mathcal{F}} .
\end{equation}
With the convention adopted above, the Maxwell contribution gives $\gamma_s=1+\mathcal{O}(\alpha)$. This is the normalization required to recover the standard weak-field refractive indices. The factor $\gamma_s$ is kept explicitly in 
the finite-field expressions and is expanded only after the convention has been fixed. This removes the apparent factor-of-two ambiguity that would arise if $\gamma_{\mathcal{F}}$ were instead taken to include the Maxwell term.

%===========================================================================
\section{Photon Dispersion and Birefringence}
\label{sec:propagation}
\label{sec:refraction}
%===========================================================================

The refractive indices of the two photon eigenmodes provide the direct link between the Heisenberg--Euler effective Lagrangian and observable vacuum birefringence. In a purely magnetic background, the external field defines a preferred spatial direction, so that photon propagation depends on both the polarization mode and the angle $\theta$ between $\mathbf{B}$ and $\mathbf{k}$.

With the normalization introduced in Eq.~(\ref{eq:gamma_s_def}), the parallel-mode refractive index for $\theta=\pi/2$ can be written as
\begin{equation}
\label{eq:n_perp_gamma}
n_{\parallel}
=
\left(
1+\frac{B^2\gamma_{\mathcal{GG}}}{\gamma_s}
\right)^{1/2}.
\end{equation}
To leading order in the one-loop correction, this gives
\begin{equation}
\label{eq:n_perp_gamma_linear}
n_{\parallel}-1
=
\frac{B^2\gamma_{\mathcal{GG}}}{2\gamma_s}
+
\mathcal{O}(\alpha^2),
\qquad
(\theta=\pi/2).
\end{equation}
For a general propagation angle, the leading birefringent correction is multiplied by $\sin^2\theta$, so that the response vanishes for propagation parallel to the external magnetic field. We define
\begin{equation}
\label{eq:xi_def}
\xi\equiv \frac{B}{B_{\rm cr}}=\frac{1}{2h}.
\end{equation}

\subsection{Refractive Indices}
\label{subsec:refractive_indices}

The corrected normalization can be checked explicitly in the weak-field limit. Since $\gamma_s=1+\mathcal{O}(\alpha)$, the leading weak-field expansions of the finite-field coefficients are
\begin{align}
\frac{B^2\gamma_{\mathcal{GG}}}{\gamma_s}
&=
\frac{\alpha}{2\pi}
\frac{14}{45}
\xi^2
+
\mathcal{O}(\xi^4,\alpha^2),
\label{eq:weak_gammaGG_check}
\\
\frac{B^2\gamma_{\mathcal{FF}}}{\gamma_s}
&=
\frac{\alpha}{2\pi}
\frac{8}{45}
\xi^2
+
\mathcal{O}(\xi^4,\alpha^2).
\label{eq:weak_gammaFF_check}
\end{align}
Substitution into the leading refractive-index relations reproduces the standard weak-field QED coefficients. This check fixes the convention used for $\gamma_s$ and removes the possible factor-of-two ambiguity.

For $\xi<1$, the weak-field refractive indices may be written as
\cite{hey97a,hey97b,hey97c}
\begin{align}
n_{\parallel}
&=
1+
\frac{\alpha}{4\pi}
\frac{14}{45}
\xi^2
\sin^2\theta
+
\mathcal{O}(\xi^4,\alpha^2),
\label{eq:n_perp_weak}
\\
n_{\perp}
&=
1+
\frac{\alpha}{4\pi}
\frac{8}{45}
\xi^2
\sin^2\theta
+
\mathcal{O}(\xi^4,\alpha^2).
\label{eq:n_parallel_weak}
\end{align}
The leading weak-field birefringence is therefore
\begin{align}
\Delta n
&\equiv
n_{\parallel}-n_{\perp}
\nonumber\\
&=
\frac{\alpha}{4\pi}
\frac{2}{15}
\xi^2
\sin^2\theta
+
\mathcal{O}(\xi^4,\alpha^2).
\label{eq:delta_n_weak}
\end{align}
These expressions show explicitly that the birefringent response vanishes for propagation parallel to the external magnetic field, $\theta=0$, and is maximal for perpendicular propagation, $\theta=\pi/2$.

For fields approaching and exceeding $B_{\rm cr}$, the weak-field expansion alone is not sufficient. The finite-field parallel-mode refractive index is computed from Eq.~(\ref{eq:n_perp_gamma}) using the one-loop expressions for $\gamma_{\mathcal{GG}}$ and $\gamma_s$. For comparison with the strong-field limit, it is useful to write the leading one-loop result in the compact form
\begin{equation}
\label{eq:n_perp_strong_compact}
n_{\parallel}
=
1+
\frac{\alpha}{4\pi}
\sin^2\theta\,
\mathcal{N}_{\parallel}(\xi)
+
\mathcal{O}(\alpha^2),
\end{equation}
where $\mathcal{N}_{\parallel}(\xi)$ denotes the corresponding finite-field coefficient. Its strong-field expansion begins as
\begin{align}
\mathcal{N}_{\parallel}(\xi)
&=
\frac{2}{3}\xi
-
C_0
-
C_1(\xi)\xi^{-1}
-
C_2\xi^{-2}
+
\mathcal{O}(\xi^{-3}).
\label{eq:N_perp_strong}
\end{align}
The coefficients are
\begin{subequations}
\label{eq:C_coeffs}
\begin{align}
C_0
&=
8\ln A
-\frac{1}{3}
-\frac{2}{3}\gamma_E,
\\
C_1(\xi)
&=
\ln\pi
+\frac{\pi^2}{18}
-2
-\ln\xi,
\\
C_2
&=
-\frac{1}{2}
-\frac{1}{6}\zeta(3).
\end{align}
\end{subequations}
Here $\gamma_E\simeq0.5772$ is the Euler--Mascheroni constant, $A\simeq1.28242712$ is the Glaisher--Kinkelin constant, and $\zeta(3)\simeq1.202$ is the Riemann zeta value \cite{olv10}. Equation~(\ref{eq:N_perp_strong}) is used only as the strong-field asymptotic form of the finite-field result, not as a replacement for the finite-field expression in the intermediate region.

For the perpendicular polarization mode, we use the one-loop expression \cite{tsa75}
\begin{equation}
\label{eq:n_parallel_compact}
n_{\perp}
=
1+
\frac{\alpha}{4\pi}
\sin^2\theta\,
\mathcal{N}_{\perp}(h)
+
\mathcal{O}(\alpha^2),
\end{equation}
where
\begin{align}
\mathcal{N}_{\perp}(h)
&=
\frac{2}{3}
+4h^2\,\psi(1+h)
-2h
-4h^2
\nonumber\\
&\quad
-4h\ln\Gamma(h)
+2h\ln(2\pi)
-2h\ln h ,
\label{eq:N_parallel}
\end{align}
so that $(\alpha/4\pi)\,\mathcal{N}_{\perp}(h)=B^2\gamma_{\mathcal{FF}}/2$ at leading one-loop order, with the weak-field limit $\mathcal{N}_{\perp}\to(8/45)\,\xi^2$ recovering Eq.~(\ref{eq:n_perp_weak}). This expression is valid for $B\leq(\pi/\alpha)B_{\rm cr}$ within the one-loop low-energy framework. We note for clarity that the corresponding exact expression for the parallel mode is Eq.~(\ref{eq:n_perp_gamma}) itself, built from 
$\gamma_{\mathcal{GG}}$; the two modes are thus controlled by the two independent second derivatives of the effective Lagrangian, and no expression is used for both.

The combined effect of a strong magnetic field and a weaker co-aligned electric field on vacuum birefringence, including rotation of the polarization vectors, has been analyzed by Kim \& Kim \cite{kim22}. Their closed-form one-loop effective Lagrangian for the combined electromagnetic wrench provides a useful generalization of the purely magnetic configuration considered here.

In the numerical analysis below, the weak-field expressions are used to check the small-$\xi$ limit, while the strong-field expansion is used only as an asymptotic comparison at large $\xi$. The transition region around $B\sim B_{\rm cr}$ is treated using the finite-field expressions rather than by extrapolating either limiting expansion.

\subsection{Birefringence and Ellipticity}
\label{subsec:birefringence_ellipticity}
\label{sec:ellipticity}

The vacuum magnetic birefringence is characterized by the difference between the two polarization-dependent refractive indices,
\begin{equation}
\label{eq:birefringence_def}
\Delta n
\equiv
n_{\parallel}-n_{\perp}.
\end{equation}
Using the compact notation introduced above, the finite-field birefringence may be written to leading order in the one-loop correction as
\begin{equation}
\label{eq:birefringence_compact}
\Delta n
=
\frac{\alpha}{4\pi}
\sin^2\theta
\left[
\mathcal{N}_{\parallel}(\xi)
-
\mathcal{N}_{\perp}(h)
\right]
+
\mathcal{O}(\alpha^2),
\end{equation}
where $\xi=B/B_{\rm cr}=1/(2h)$. In the weak-field limit, 
Eq.~(\ref{eq:birefringence_compact}) reduces to the standard Cotton--Mouton form,
\begin{equation}
\label{eq:cotton_mouton}
\Delta n
=
k_{\rm CM} B^2 \sin^2\theta,
\end{equation}
where the QED Cotton--Mouton coefficient is
\begin{equation}
\label{eq:kCM}
k_{\rm CM}
=
\frac{\alpha}{30\pi}
\frac{1}{B_{\rm cr}^2}
\simeq
4.0\times10^{-24}\,{\rm T}^{-2}.
\end{equation}
Equivalently,
\begin{equation}
\label{eq:delta_n_weak_again}
\Delta n
=
\frac{\alpha}{30\pi}
\left(
\frac{B}{B_{\rm cr}}
\right)^2
\sin^2\theta
+
\mathcal{O}
\left[
\left(
\frac{B}{B_{\rm cr}}
\right)^4,
\alpha^2
\right].
\end{equation}
This expression makes explicit that magnetic vacuum birefringence vanishes for propagation parallel to the external field and is maximal for propagation perpendicular to it.

An important observable consequence of vacuum birefringence is the ellipticity $\chi$ acquired by a linearly polarized electromagnetic wave after propagating through a magnetized region. It is defined by
\begin{equation}
\chi
=
\frac{1}{2}k\,(n_{\parallel}-n_{\perp})\,\ell
=
\frac{\pi}{\lambda}\,\Delta n\,\ell,
\label{eq:ellipticity_def}
\end{equation}
where $k=|\mathbf{k}|=2\pi/\lambda$ is the magnitude of the photon wave vector, $\lambda$ is the wavelength, and $\ell$ is the path length in the external magnetic field region. Equation~(\ref{eq:ellipticity_def}) applies in the regime of small phase retardation,
\begin{equation}
|\Delta n|\,k\,\ell\ll1,
\end{equation}
for which the induced ellipticity remains a useful linear observable. For larger phase shifts, a full Jones-matrix or Stokes-parameter treatment is more appropriate. In extended astrophysical magnetospheres, the accumulated phase retardation may become non-negligible because even a small refractive-index difference can be amplified by a large propagation length; however, quantitative predictions require a full polarization-transport calculation.

In the weak-field limit, using Eq.~(\ref{eq:delta_n_weak_again}), one obtains
\begin{equation}
\chi
\simeq
\frac{\alpha}{60\pi}
\left(
\frac{B}{B_{\rm cr}}
\right)^2
\frac{\omega\ell}{c}
\sin^2\theta,
\label{eq:ellipticity_weak}
\end{equation}
where $\omega=2\pi c/\lambda$. Thus, the ellipticity is quadratic in the external magnetic field strength and increases linearly with optical path length and photon frequency.

PVLAS provides the most direct laboratory context for this observable. The experiment probes ellipticity in a magnetic field of order $B\simeq2.5$~T using light of wavelength $\lambda\simeq1064$~nm and an effective optical path length 
$\ell_{\rm eff}\simeq36$~km, obtained through a Fabry--P\'erot cavity with approximately $4.4\times10^{4}$ passes. For these parameters, the QED prediction corresponds to $\Delta n_{\rm QED}\simeq2.5\times10^{-23}$ and an ellipticity of order $\chi_{\rm QED}\sim10^{-11}$--$10^{-12}$~rad. The current PVLAS sensitivity \cite{pvlas20} is therefore approaching the range relevant for a direct laboratory test of magnetic vacuum birefringence.

\begin{figure*}[t]
\centering
\includegraphics[width=\textwidth]{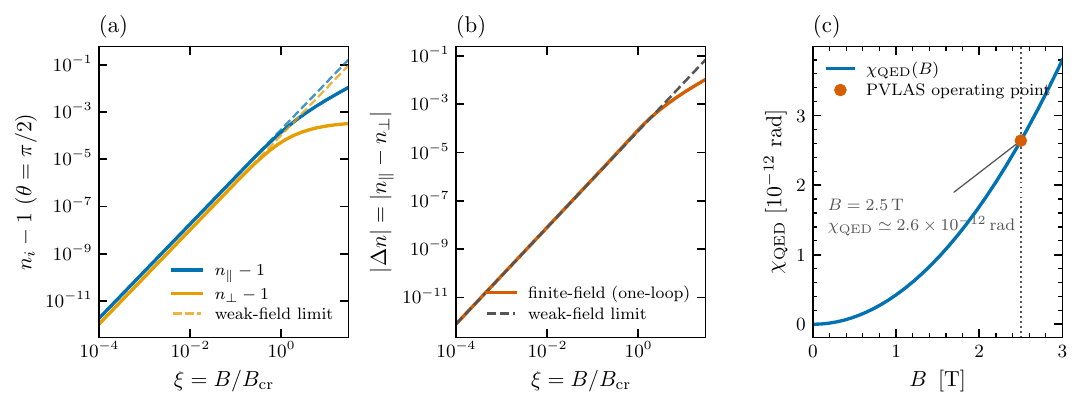}
\caption{\textbf{Refractive index, vacuum birefringence, and laboratory ellipticity, 
computed from the exact one-loop finite-field expressions.} \textbf{Panel (a):} 
Refractive-index deviations $n_\parallel-1$ and $n_\perp-1$ at $\theta=\pi/2$ versus 
$\xi=B/B_{\rm cr}$ (solid), compared with the weak-field $\xi^2$ scaling (dashed). 
\textbf{Panel (b):} Vacuum birefringence $|\Delta n|=|n_\parallel-n_\perp|$ (solid), 
compared with the weak-field Cotton--Mouton form of Eq.~(\ref{eq:cotton_mouton}) 
(dashed). \textbf{Panel (c):} PVLAS-parameter ellipticity 
$\chi_{\rm QED}(B)$ of Eq.~(\ref{eq:ellipticity_weak}), for 
$\lambda=1064$~nm and $\ell_{\rm eff}=36$~km, with the PVLAS operating point 
$B=2.5$~T marked.}
\label{fig:refractive}
\end{figure*}

Figure~\ref{fig:refractive} illustrates these results numerically, using the exact one-loop expressions of Eqs.~(\ref{eq:gammaF})--(\ref{eq:gammaGG}) rather than the truncated weak-field series alone. Panel (a) shows the refractive-index 
deviations $n_\parallel-1$ and $n_\perp-1$ as a function of $\xi=B/B_{\rm cr}$; both curves track the weak-field $\xi^2$ scaling for $\xi\lesssim1$ and depart from it as $\xi\to30$. Panel (b) shows the resulting birefringence $|\Delta n|$, again compared with the weak-field Cotton--Mouton scaling of Eq.~(\ref{eq:cotton_mouton}). Panel (c) evaluates the PVLAS-parameter ellipticity of Eq.~(\ref{eq:ellipticity_weak}) directly as a function of laboratory field strength and confirms the order-of-magnitude estimate $\chi_{\rm QED}(2.5\,{\rm T})\simeq2.6\times10^{-12}$~rad quoted above.

%===========================================================================
\section{Photon Magnetic Response}
\label{sec:mu}
%===========================================================================

The magnetic response of a photon in a magnetized QED vacuum is defined through the field dependence of its polarization-dependent dispersion relation. For the two physical photon eigenmodes,
\begin{equation}
\label{eq:omega_modes_mu}
\omega_i(B,\theta)
=
\frac{|\mathbf{k}|}{n_i(B,\theta)},
\qquad
i=\parallel,\perp .
\end{equation}
Here $n_i(B,\theta)$ is the refractive index of the corresponding polarization mode, and $\theta$ is the angle between the external magnetic field $\mathbf{B}$ and the photon wave vector $\mathbf{k}$.

The refractive indices can be written in terms of the finite-field quantities
\begin{equation}
\label{eq:kappa_defs}
\kappa_s
=
\frac{B^2\gamma_{\mathcal{FF}}}{\gamma_s},
\qquad
\kappa_p
=
\frac{B^2\gamma_{\mathcal{GG}}}{\gamma_s},
\qquad
\gamma_s=1-\gamma_{\mathcal{F}} .
\end{equation}
With the normalization convention fixed in Eq.~(\ref{eq:gamma_s_def}), the factor $\gamma_s$ is kept explicitly. This is essential because the magnetic response involves a derivative with respect to $B$, and replacing $\gamma_s$ by a fixed number too early can introduce a normalization ambiguity.

For the two photon eigenmodes, the corresponding refractive indices may be written as
\cite{cha12}
\begin{subequations}
\label{eq:n_kappa_modes}
\begin{align}
n_{\perp}
&=
\frac{1}{\sqrt{1-\kappa_s\sin^2\theta}},
\\
n_{\parallel}
&=
\left(
\frac{1+\kappa_p}{1+\kappa_p\cos^2\theta}
\right)^{1/2}.
\end{align}
\end{subequations}
Equations~(\ref{eq:n_kappa_modes}) follow from the uniaxial vacuum permittivities of Ref.~\cite{cha12}, $\varepsilon_\perp=\mu_\perp^{-1}=\gamma_s$, $\varepsilon_\parallel=\gamma_s(1+\kappa_p)$, and $\mu_\parallel^{-1}=\gamma_s(1-\kappa_s)$, applied to the two eigenmodes of a uniaxial medium. Both indices reduce to unity for propagation parallel to the external field, $\theta=0$, and to $1+\tfrac12\kappa\sin^2\theta$ at leading 
one-loop order, consistent with the $\sin^2\theta$ dependence of Eqs.~(\ref{eq:n_perp_weak})--(\ref{eq:n_parallel_weak}). All finite-field quantities below are evaluated at $\theta=\pi/2$. For perpendicular propagation, $\theta=\pi/2$, Eq.~(\ref{eq:n_kappa_modes}) reduces to
\begin{equation}
\label{eq:n_perp_90_mu}
n_{\parallel}
=
\sqrt{1+\kappa_p}.
\end{equation}

\subsection{Definition and Finite-Field Form}

The effective magnetic moment of a photon eigenmode is defined by
\begin{equation}
\label{eq:mu_def}
\mu_{\gamma}^{(i)}
=
-
\left(
\frac{\partial \omega_i}{\partial B}
\right)_{|\mathbf{k}|,\theta},
\qquad
i=\parallel,\perp .
\end{equation}
Equivalently, using Eq.~(\ref{eq:omega_modes_mu}),
\begin{equation}
\label{eq:mu_from_n}
\mu_{\gamma}^{(i)}
=
\frac{|\mathbf{k}|}{n_i^2}
\left(
\frac{\partial n_i}{\partial B}
\right)_{|\mathbf{k}|,\theta}.
\end{equation}
This definition gives a scalar magnetic response associated with each photon eigenmode. The angular dependence enters through the mode-dependent refractive indices; no independent transverse magnetic-moment component is introduced.

For the parallel mode at $\theta=\pi/2$, Eqs.~(\ref{eq:n_perp_90_mu}) and 
(\ref{eq:mu_from_n}) give
\begin{equation}
\label{eq:mu_perp_exact_kappa}
\mu_{\gamma}^{(\parallel)}
=
\frac{|\mathbf{k}|}{2(1+\kappa_p)^{3/2}}
\frac{d\kappa_p}{dB}.
\end{equation}
Using Eq.~(\ref{eq:kappa_defs}), this becomes
\begin{equation}
\label{eq:mu_perp_exact_gamma}
\mu_{\gamma}^{(\parallel)}
=
\frac{|\mathbf{k}|}{2(1+\kappa_p)^{3/2}}
\frac{d}{dB}
\left(
\frac{B^2\gamma_{\mathcal{GG}}}{\gamma_s}
\right).
\end{equation}
Equation~(\ref{eq:mu_perp_exact_gamma}) is the compact finite-field expression used for the parallel-mode photon magnetic response. The derivative acts on the full ratio $B^2\gamma_{\mathcal{GG}}/\gamma_s$, not on$B^2\gamma_{\mathcal{GG}}$ alone.

To leading order in the one-loop correction,
\begin{equation}
\label{eq:mu_perp_leading}
\mu_{\gamma}^{(\parallel)}
=
\frac{|\mathbf{k}|}{2}
\frac{d}{dB}
\left(
\frac{B^2\gamma_{\mathcal{GG}}}{\gamma_s}
\right)
+
\mathcal{O}(\alpha^2).
\end{equation}
In terms of the dimensionless magnetic field
\begin{equation}
\label{eq:xi_mu_def}
\xi
=
\frac{B}{B_{\rm cr}},
\qquad
B_{\rm cr}
=
\frac{m^2}{e},
\end{equation}
the magnetic response can be written as
\begin{equation}
\label{eq:mu_dimensionless}
\mu_{\gamma}^{(i)}
=
\frac{e}{m}
\frac{|\mathbf{k}|}{m}
\,
\mathcal{M}_{i}(\xi,\theta),
\end{equation}
where $\mathcal{M}_{i}$ is a dimensionless response function obtained by differentiating $n_i$ with respect to $\xi$. This form makes explicit the low-energy suppression factor $|\mathbf{k}|/m$ relative to the natural magnetic moment scale $e/m$.

The same response can also be expressed through the phase velocity. For fixed propagation direction,
\begin{equation}
\label{eq:v_phase_i}
v_{{\rm ph},i}
=
\frac{\omega_i}{|\mathbf{k}|}
=
\frac{1}{n_i(B,\theta)} .
\end{equation}
Therefore,
\begin{equation}
\label{eq:mu_v_relation}
\mu_{\gamma}^{(i)}
=
-
|\mathbf{k}|
\left(
\frac{\partial v_{{\rm ph},i}}{\partial B}
\right)_\theta .
\end{equation}
Thus, a positive photon magnetic response corresponds to a decrease of the phase velocity as the external magnetic field is increased. For the parallel mode,
\begin{equation}
\label{eq:v_perp_exact}
v_{{\rm ph},\parallel}^{\,2}
=
\frac{1}{n_\parallel^2}
=
\frac{1}{1+\kappa_p},
\end{equation}
and, to leading order in the one-loop correction,
\begin{equation}
\label{eq:v_perp_leading}
v_{{\rm ph},\parallel}^{\,2}
=
1-\kappa_p
+
\mathcal{O}(\alpha^2).
\end{equation}
This shows that the same finite-field quantity controlling the parallel-mode refractive index also controls the reduction of the photon phase velocity in the magnetized vacuum.

\subsection{Weak- and Strong-Field Limits}

Using the weak-field refractive indices in Eqs.~(\ref{eq:n_perp_weak}) and (\ref{eq:n_parallel_weak}), the leading magnetic responses of the two polarization 
modes are
\begin{subequations}
\label{eq:mu_weak_modes}
\begin{align}
\mu_{\gamma}^{(\parallel)}
&=
\frac{e}{m}
\frac{|\mathbf{k}|}{m}
\frac{\alpha}{4\pi}
\frac{28}{45}
\xi
\sin^2\theta
+
\mathcal{O}(\xi^3,\alpha^2),
\\
\mu_{\gamma}^{(\perp)}
&=
\frac{e}{m}
\frac{|\mathbf{k}|}{m}
\frac{\alpha}{4\pi}
\frac{16}{45}
\xi
\sin^2\theta
+
\mathcal{O}(\xi^3,\alpha^2).
\end{align}
\end{subequations}
Thus the weak-field magnetic response is linear in $B$, proportional to the photon energy $|\mathbf{k}|$, and vanishes for propagation parallel to the external field. The coefficient $28/45$ in the parallel mode follows directly from 
differentiating the leading weak-field refractive-index correction proportional to $(B/B_{\rm cr})^2$; it is not a contribution from both polarization modes.

The weak-field birefringent part of the magnetic response is
\begin{equation}
\label{eq:delta_mu_weak}
\mu_{\gamma}^{(\parallel)}
-
\mu_{\gamma}^{(\perp)}
=
\frac{e}{m}
\frac{|\mathbf{k}|}{m}
\frac{\alpha}{4\pi}
\frac{4}{15}
\xi
\sin^2\theta
+
\mathcal{O}(\xi^3,\alpha^2).
\end{equation}
This quantity measures the magnetic-field derivative of the polarization splitting and is the magnetic-response analogue of vacuum birefringence.

For $B\gtrsim B_{\rm cr}$, the parallel-mode response approaches its asymptotic one-loop behavior. In terms of $\xi=B/B_{\rm cr}$,
\begin{align}
\mu_{\gamma}^{(\parallel)}
&=
\frac{e}{m}
\frac{|\mathbf{k}|}{m}
\frac{\alpha}{4\pi}
\sin^2\theta
\notag\\
&\quad\times
\left[
\frac{2}{3}
+
\frac{
\ln\pi+\pi^2/18-1-\ln\xi
}{\xi^2}
\right.
\notag\\
&\hspace{2.25cm}\left.
+
\mathcal{O}(\xi^{-3})
\right].
\label{eq:mu_strong_compact}
\end{align}
The corresponding strong-field asymptotic value is
\begin{equation}
\label{eq:mu_asymptotic}
\mu_{\gamma}^{(\parallel)}
\rightarrow
\frac{e}{m}
\frac{|\mathbf{k}|}{m}
\frac{\alpha}{4\pi}
\frac{2}{3}
\sin^2\theta,
\qquad
\xi\gg1 .
\end{equation}
For photons satisfying $|\mathbf{k}|/m\ll1$, this remains much smaller than the natural magnetic-moment scale $e/m$, as required by the low-energy approximation.

\subsection{Domain of Validity}
\label{sec:validity}

The expressions above are derived within the one-loop Heisenberg--Euler framework. 
They require
\begin{equation}
\label{eq:validity}
\frac{|\mathbf{k}|}{m}\ll1,
\end{equation}
so that the photon energy is well below the pair-creation threshold and derivative corrections to the effective action can be neglected. The background field is also assumed to be constant or slowly varying on the electron Compton scale. The formal one-loop expressions can be used to explore the range $0\leq B\leq30\,B_{\rm cr}$, but higher-loop corrections and pair-production effects must be assessed separately when the field strength or photon energy approaches the 
limits of the low-energy approximation.

The magnetic response is paramagnetic when
\begin{equation}
\label{eq:paramagnetic}
\mu_{\gamma}^{(i)}(B,\theta)>0 .
\end{equation}
For the parallel mode, this is equivalent to the statement that the photon energy decreases as the external magnetic field increases. In the weak-field limit, positivity follows immediately from Eq.~(\ref{eq:mu_weak_modes}). In the finite-field regime, positivity and monotonicity are tested using the derivative form in Eq.~(\ref{eq:mu_perp_exact_gamma}). Evaluated with the exact one-loop special-function expressions of Eqs.~(\ref{eq:gammaF})--(\ref{eq:gammaGG}), the 
response $\widehat{\mu}_\gamma^{(\parallel)}$ is strictly positive over the entire interval $0\leq B\leq30\,B_{\rm cr}$ and increases monotonically up to a broad maximum near $B\simeq17\,B_{\rm cr}$; beyond this point it decreases slightly, by less than $1\%$, toward $B=30\,B_{\rm cr}$ (see Fig.~\ref{fig:experimental}, panel b). The response is therefore not monotonic over the full stated interval in the strict mathematical 
sense, although the non-monotonic part of the interval represents a sub-percent-level correction and the response remains within a few percent of its strong-field asymptotic value throughout. As discussed in Sec.~\ref{sec:peak_analytic}, 
this non-monotonicity is a property of the resummed normalization $(1+\kappa_p)^{-3/2}$: the strictly truncated $\mathcal{O}(\alpha)$ response $\tfrac12\kappa_p'$ is monotonic, and the maximum is order-controlled only in its 
existence and approximate location, not in its sub-percent profile. These tests should be regarded as numerical validation over the field interval studied, not as a global analytic proof unless the required inequalities are established explicitly.

For comparison, the anomalous magnetic moment of the electron is
\begin{equation}
\label{eq:mu_electron}
\mu_{\rm e,anom}
=
\frac{\alpha}{2\pi}
\frac{1}{2}
\frac{e}{m}.
\end{equation}
Using Eq.~(\ref{eq:mu_asymptotic}), one obtains
\begin{equation}
\label{eq:mu_comparison}
\mu_{\gamma}^{(\parallel)}
\rightarrow
\frac{|\mathbf{k}|}{m}
\frac{2}{3}
\mu_{\rm e,anom}
\sin^2\theta .
\end{equation}
Thus, for photons satisfying $|\mathbf{k}|/m\ll1$, the effective photon magnetic moment is suppressed relative to the electron anomalous magnetic moment by the low-energy factor $|\mathbf{k}|/m$. Only in the formal limit $|\mathbf{k}|\sim m$ does the asymptotic coefficient become comparable to $(2/3)\mu_{\rm e,anom}$; this limit lies at the edge of the low-energy regime and should not be overinterpreted.

The photon magnetic moment considered here is therefore not an intrinsic magnetic moment of a free photon. It is an effective magnetic response of the photon dispersion relation induced by vacuum polarization in an external magnetic field. The magnetized vacuum behaves as an anisotropic optical medium, and the response depends on both the polarization eigenmode and the propagation angle relative to the external field.

%===========================================================================
\section{Numerical Results and Applications}
\label{sec:numerics_context}
\label{sec:experiment}
\label{sec:observables}
%===========================================================================

The finite-field magnetic response derived above is not measured directly in current experiments. Instead, it is connected to observable phenomena through the same polarization-dependent refractive indices that generate vacuum birefringence, ellipticity, polarization transport, and photon--photon scattering. In this section, we first summarize the numerical validation of the finite-field response and then place the result in its laboratory and astrophysical context.

\subsection{Numerical Evaluation}

To test the analytic expressions derived above, we numerically evaluate the finite-field magnetic response of the photon eigenmodes over the range $0\leq B\leq30\,B_{\rm cr}$. We focus on the parallel mode at $\theta=\pi/2$, where the magnetic response is maximal and where comparison with the weak- and strong-field limits is most transparent. The numerical analysis is based on the derivative form
\begin{equation}
\label{eq:mu_numerical_definition}
\mu_{\gamma}^{(\parallel)}
=
\frac{|\mathbf{k}|}{2(1+\kappa_p)^{3/2}}
\frac{d}{dB}
\left(
\frac{B^2\gamma_{\mathcal{GG}}}{\gamma_s}
\right),
\end{equation}
with $\gamma_s$ kept explicitly. The derivative acts on the full ratio $B^2\gamma_{\mathcal{GG}}/\gamma_s$, rather than on $B^2\gamma_{\mathcal{GG}}$ alone. This is important because replacing $\gamma_s$ by a fixed number before 
differentiation can introduce a normalization ambiguity in the inferred magnetic response.

The normalized response used below is
\begin{equation}
\label{eq:mu_normalized}
\widehat{\mu}_{\gamma}^{(\parallel)}
\equiv
\frac{m}{e}
\frac{m}{|\mathbf{k}|}
\mu_{\gamma}^{(\parallel)} .
\end{equation}
With this normalization, the strong-field asymptotic value for $\theta=\pi/2$ is
\begin{equation}
\label{eq:mu_normalized_asymptotic}
\widehat{\mu}_{\gamma}^{(\parallel)}
\rightarrow
\frac{\alpha}{4\pi}
\frac{2}{3},
\qquad
B\gg B_{\rm cr}.
\end{equation}
The numerical calculation is checked against four criteria: agreement with the weak-field expansion of Eq.~(\ref{eq:mu_weak_modes}) at small $\xi$, approach to the strong-field form of Eq.~(\ref{eq:mu_asymptotic}) at large $\xi$, positivity, and monotonicity. The first three are satisfied to high precision: the finite-field result agrees with the weak-field expansion to better than $0.02\%$ for $\xi\lesssim10^{-2}$ and reaches $96$--$97\%$ of the strong-field asymptotic value by $\xi=30$. The fourth requires the refined statement already given in Sec.~\ref{sec:validity}: monotonic growth up to a broad maximum at $\xi\simeq17$, followed by a sub-percent decrease out to $\xi=30$ -- a shape that belongs to the resummed normalization and whose perturbative status is assessed in Sec.~\ref{sec:peak_analytic}. These checks validate the finite-field formulation over the stated interval; they do not constitute a global analytic proof outside it.

\subsection{Extremum of the Resummed Response}
\label{sec:peak_analytic}

\begin{figure*}[t]
\centering
\includegraphics[width=\textwidth]{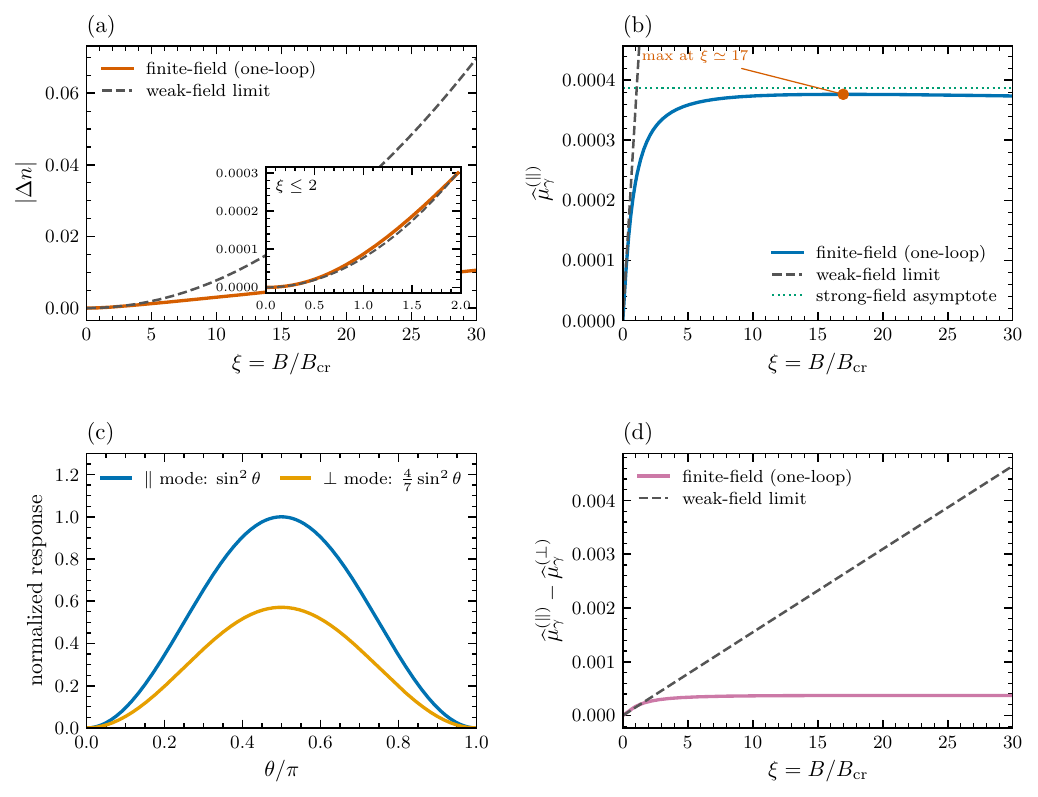}
\caption{\textbf{Numerical behavior of the finite-field photon magnetic response.} 
\textbf{Panel (a):} Vacuum magnetic birefringence $|\Delta n|$ as a function of 
$B/B_{\rm cr}$ over the studied range $0\leq\xi\leq30$ (inset: weak-field close-up, 
$\xi\leq2$), showing the weak-field scaling and the transition toward the 
strong-field regime. 
\textbf{Panel (b):} Normalized parallel-mode magnetic response 
$\widehat{\mu}_{\gamma}^{(\parallel)}$ computed from 
Eq.~(\ref{eq:mu_numerical_definition}), compared with the weak-field expansion and 
the leading-order strong-field asymptote. The response is positive 
throughout and increases monotonically up to a broad maximum near $\xi\simeq17$ 
(marked), after which it decreases by less than $1\%$ toward $\xi=30$; it does not 
approach the strong-field asymptote monotonically over the full interval shown. 
This shape belongs to the resummed one-loop normalization; the strictly 
truncated $\mathcal{O}(\alpha)$ response is monotonic (Sec.~\ref{sec:peak_analytic} 
and Fig.~\ref{fig:peak}). 
\textbf{Panel (c):} Angular ($\sin^2\theta$) dependence of the weak-field magnetic 
response for both eigenmodes, normalized to $\widehat{\mu}_{\gamma}^{(\parallel)}(\pi/2)$, 
illustrating the vanishing response for propagation parallel to the external field. 
\textbf{Panel (d):} Difference between the perpendicular and parallel magnetic 
responses, $\widehat{\mu}_{\gamma}^{(\parallel)}-\widehat{\mu}_{\gamma}^{(\perp)}$, 
showing the magnetic-response analogue of vacuum birefringence.}
\label{fig:experimental}
\end{figure*}

The existence of a maximum in $\widehat{\mu}_{\gamma}^{(\parallel)}(\xi)$ is not an unexplained numerical feature: it can be located analytically. Writing $\widehat{\mu}_{\gamma}^{(\parallel)}(\xi)=\tfrac12(1+\kappa_p)^{-3/2}\kappa_p'(\xi)$ (Eq.~\ref{eq:mu_perp_exact_gamma}) and setting its derivative with respect to $\xi$ to zero gives the general extremum condition
\begin{equation}
\label{eq:peak_condition_main}
(1+\kappa_p)\,\kappa_p''=\frac32\,(\kappa_p')^2 ,
\end{equation}
derived and discussed in detail in Appendix~\ref{app:positivity}. This condition is not specific to the parallel-mode $\gamma_{\mathcal{GG}}$ used here; it applies to any resummed response of the form $n=\sqrt{1+\kappa(\xi)}$ together with the physical normalization $\mu\propto n^{-2}n'$, and identifies a maximum whenever $\kappa'(\xi)$ approaches a bounded value while $\kappa(\xi)$ itself grows without bound -- precisely the situation established for $\kappa_p$ in Appendix~\ref{app:positivity}. Solving Eq.~(\ref{eq:peak_condition_main}) numerically using the exact special-function expressions of Eqs.~(\ref{eq:gammaF})--(\ref{eq:gammaGG}) reproduces the maximum found in Fig.~\ref{fig:experimental}, panel (b), to full numerical precision: $\xi_{\rm peak}=16.963$.

A closed-form estimate of $\xi_{\rm peak}$ can be obtained without any new numerical machinery, using only the strong-field expansion already derived in Sec.~\ref{sec:propagation} (Eqs.~\ref{eq:N_perp_strong}--\ref{eq:C_coeffs}). 
Substituting $\kappa_p\simeq(\alpha/2\pi)\mathcal{N}_\parallel(\xi)$ into Eq.~(\ref{eq:peak_condition_main}) and solving for its root gives $\xi_{\rm peak}\simeq16.73$ (Appendix~\ref{app:positivity}, Eq.~\ref{eq:A_peak_closed}), 
in agreement with the exact value to within $1.4\%$. This agreement is a useful internal consistency check: it shows that the location of the maximum is controlled by the same strong-field coefficients, $C_0$, $C_1(\xi)$, and $C_2$, that already appear elsewhere in this work, rather than being a separate, unrelated feature of the exact special functions.

\textit{Perturbative status of the maximum.}---The interpretation of this feature requires care, because it is not order-controlled in $\alpha$. At strict one-loop accuracy, $(1+\kappa_p)^{-3/2}=1-\tfrac32\kappa_p+\mathcal{O}(\alpha^2)$, so the consistently truncated response $\tfrac12\kappa_p'(\xi)$ is \emph{monotonic} over the entire interval studied and saturates toward the strong-field asymptote from below, with no maximum [Fig.~\ref{fig:peak}, dash-dotted curve]. The maximum exists only when the one-loop constitutive relation $n_\parallel=\sqrt{1+\kappa_p}$ is solved without expansion: the resummed normalization retains a selected, geometric subset of $\mathcal{O}(\alpha^2)$ and higher terms. At the peak the retained suppression is $(1+\kappa_p)^{-3/2}-1\simeq-1.8\%$ [$\kappa_p(\xi_{\rm peak})=0.012$], while genuine two-loop Heisenberg--Euler corrections, which are omitted here, modify $\kappa_p$ at relative order $\alpha/\pi$ up to logarithms of the field strength \cite{rit75,gie17,kar19}; the two effects are of comparable formal order. The structural statement is nevertheless robust: the extremum condition of Eq.~(\ref{eq:peak_condition_main}) shows that \emph{any} correction preserving the established behavior -- $\kappa_p'$ bounded while $\kappa_p$ grows without bound, which the known two-loop and strong-field all-loop results respect \cite{gie17,kar19} -- preserves the existence of the maximum and shifts its location only at relative $\mathcal{O}(\alpha)$. We therefore quote $\xi_{\rm peak}\simeq17$ as the physical statement; the five-digit value $16.963$ is exact only within the resummed one-loop model, and the sub-percent decrease of the response beyond the peak lies below the expected size of higher-loop corrections and should not be regarded as a controlled prediction.

Figure~\ref{fig:experimental} summarizes the numerical behavior of the finite-field magnetic response. The figure is intended to emphasize the theory: the weak-field limit, the finite-field numerical result, the strong-field asymptote, and the angular or polarization dependence of $\mu_{\gamma}^{(\parallel,\perp)}$. Experimental and astrophysical systems should be shown only as contextual field-scale markers unless a full observational model is included.

\subsection{Experimental and Astrophysical Context}

PVLAS provides the most direct laboratory context for the birefringence sector of the present work. For the reported magnetic field strength $B=2.5\,{\rm T}$, the QED prediction is
\begin{equation}
\label{eq:pvlas_qed_prediction}
\Delta n_{\rm QED}
=
k_{\rm CM}B^2
\simeq
2.5\times10^{-23},
\end{equation}
using $k_{\rm CM}\simeq4.0\times10^{-24}\,{\rm T}^{-2}$. The current PVLAS result is
\begin{equation}
\label{eq:pvlas_measurement}
\Delta n_{\rm PVLAS}
=
(12\pm17)\times10^{-23}.
\end{equation}
This measurement is consistent with zero within the quoted uncertainty, but its sensitivity is approaching the scale of the QED prediction \cite{pvlas20}. PVLAS therefore provides direct experimental context for magnetic vacuum birefringence. It does not directly measure the photon magnetic moment $\mu_\gamma(B)$, but the same refractive-index splitting that defines magnetic birefringence also enters the field derivative used to define the magnetic response.

Photon--photon scattering is another manifestation of the nonlinear QED vacuum. In the low-energy limit, the total photon--photon scattering cross-section derived from the Heisenberg--Euler effective Lagrangian scales as
\begin{equation}
\label{eq:gamma_gamma}
\sigma_{\gamma\gamma\rightarrow\gamma\gamma}
\simeq
\frac{973}{10125\pi}
\alpha^4
\frac{\omega^6}{m^8},
\qquad
\omega\ll m .
\end{equation}
The ATLAS observation of light-by-light scattering in ultra-peripheral Pb--Pb collisions, with measured cross-section $\sigma_{\gamma\gamma\rightarrow\gamma\gamma}=78\pm15\,{\rm nb}$, provides a high-energy experimental confirmation of photon--photon scattering in QED \cite{atlas17,atlas19}. This result supports the nonlinear-QED framework, but it probes a different kinematic regime from the low-energy, static-field calculation considered here. It should therefore be cited as complementary evidence for nonlinear photon interactions, not as a direct test of the photon magnetic moment in an external magnetic field.

Strongly magnetized neutron stars provide astrophysical environments in which magnetic fields can approach or exceed $B_{\rm cr}$. In such systems, QED vacuum birefringence can modify the polarization transport of radiation through the magnetosphere. The degree of linear polarization may be written as
\begin{equation}
\label{eq:pol_degree}
\Pi
=
\frac{I_\perp-I_\parallel}{I_\perp+I_\parallel},
\end{equation}
where $I_\perp$ and $I_\parallel$ are the intensities associated with the two polarization modes. The refractive-index difference $\Delta n=n_\parallel-n_\perp$ derived in this work provides one microscopic QED input to such calculations. 
However, the observed polarization degree cannot be predicted from $\Delta n$ or $\mu_\gamma$ alone. A quantitative comparison with magnetar data requires radiative transfer through the magnetosphere, the surface emission model, viewing geometry, plasma effects, and magnetic-field topology.

Recent IXPE observations provide important astrophysical motivation for this problem. Examples include the polarization-angle swing in 4U~0142+61 \cite{taverna22,lai23}, polarization measurements of 1RXS~J1708-4009 \cite{zane23}, and high polarization fractions in 1E~1547.0-5408 \cite{ixpe25}. Optical polarimetry of the isolated neutron star RX~J1856.5-3754 has also provided suggestive evidence for vacuum birefringence effects \cite{mig17}. These observations are highly relevant to the physical motivation of the present work, but they should be regarded as astrophysical context rather than as direct validation of the finite-field photon magnetic moment.

The observational picture for 1E~1547.0$-$5408 has continued to develop since the original submission of this work: following up on the high polarization degree reported in Ref.~\cite{ixpe25}, a dedicated 500~ks IXPE campaign found a phase-averaged polarization degree of $47.7\pm2.9\%$ in the $2$--$6$~keV band, together with a mild, $\sim1\sigma$ dip in polarization degree between $3$ and $4$~keV that is qualitatively consistent with mode conversion at the vacuum resonance \cite{tav26}, while forecasts using realistic field profiles identify 1RXS~J170849.0$-$400910 as the most promising near-term target for vacuum-birefringence-induced phase delays with IXPE and eXTP \cite{abu26}. A quantitative, model-based confrontation between finite-field QED predictions and magnetar X-ray polarimetry is therefore becoming feasible; Sec.~\ref{sec:vres} carries out the part of that confrontation that the present calculation supports, by propagating the exact $\Delta n$ into the vacuum-resonance observables of these two sources.

\begin{figure}[t]
\centering
\includegraphics[width=0.95\columnwidth]{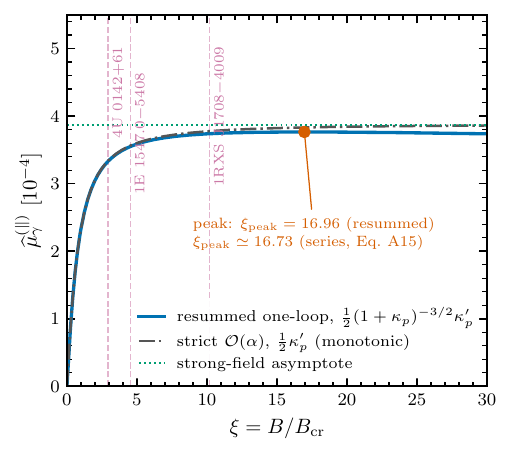}
\caption{\textbf{Analytic peak characterization and magnetar field-strength context.} 
The finite-field one-loop response $\widehat{\mu}_{\gamma}^{(\parallel)}(\xi)$ with the 
resummed normalization (solid), the strictly truncated $\mathcal{O}(\alpha)$ response 
$\tfrac12\kappa_p'(\xi)$ (dash-dotted), which is monotonic and exhibits no maximum, 
the leading-order strong-field asymptote (dotted), and the maximum located by the 
extremum condition of Eq.~(\ref{eq:A_extremum}) (marker; exact value and closed-form 
estimate from Eq.~\ref{eq:A_peak_closed} both quoted). The $\simeq2\%$ splitting 
between the two response curves at the peak is the resummation effect discussed in 
Sec.~\ref{sec:peak_analytic}; it is of the same formal order as omitted two-loop 
corrections \cite{rit75,gie17,kar19}. Vertical dashed lines mark the field 
strengths, expressed as $\xi=B/B_{\rm cr}$, of the magnetars already discussed in the 
text: 4U~0142+61, 1E~1547.0$-$5408, and 1RXS~J1708$-$4009 \cite{kas14}.}
\label{fig:peak}
\end{figure}

A more specific connection can be drawn between the analytic result of Sec.~\ref{sec:peak_analytic} and the magnetars already discussed above. The maximum of $\widehat{\mu}_{\gamma}^{(\parallel)}(\xi)$ identified there occurs at $B_{\rm peak}=\xi_{\rm peak}B_{\rm cr}\simeq7.5\times10^{14}$~G. This field strength is comparable to the spin-down-inferred surface dipole fields of the magnetars already considered in this work: $B\simeq1.3\times10^{14}$~G for 4U~0142+61 ($\xi\simeq2.9$), $B\simeq2\times10^{14}$~G for 1E~1547.0$-$5408 ($\xi\simeq4.5$), and $B\simeq4$--$5\times10^{14}$~G for 1RXS~J1708$-$4009 ($\xi\simeq10$--$11$) \cite{kas14}; more strongly magnetized sources, with fields approaching or exceeding $10^{15}$~G, correspond to $\xi\gtrsim20$, beyond the peak. Figure~\ref{fig:peak} shows the finite-field response together with these field strengths marked on the same axis. The known magnetar population therefore straddles the theoretically identified transition scale rather than lying entirely to one side of it. This is a suggestive connection, not yet a testable prediction: what magnetar X-ray polarimetry measures is not $\widehat{\mu}_{\gamma}^{(\parallel)}$ directly but the accumulated effect of $\Delta n$ on radiative transfer through a magnetosphere in which $B$ varies continuously from the stellar surface outward, so a quantitative comparison would require folding the finite-field response derived here into a full polarization-transport calculation of the kind already noted above to lie beyond the present scope. Nonetheless, the coincidence indicates that the non-monotonic feature identified in Sec.~\ref{sec:peak_analytic} occurs in a field-strength regime of direct astrophysical relevance, rather than being confined to a purely formal extrapolation of the theory.

Several experimental and observational programs are relevant for future tests of nonlinear QED. Laboratory efforts such as PVLAS and BMV continue to improve polarimetric sensitivity in magnetic fields \cite{pvlas20,bre10,cad14}. X-ray free-electron-laser facilities, including HIBEF at the European XFEL, offer possible routes to vacuum birefringence measurements using high-intensity laser and X-ray polarimetry configurations \cite{hibef24}. In astrophysics, continued X-ray polarimetry of neutron stars and magnetars will further test polarization transport in ultra-strong magnetic fields. These directions motivate the need for accurate finite-field theoretical inputs. The present calculation contributes to this program by providing a controlled description of the magnetic response of photon eigenmodes in the one-loop Heisenberg--Euler framework.

\subsection{Polarization-Limiting Radius}
\label{sec:rpl}

The preceding subsections used magnetar polarimetry as context; here we make the connection quantitative for one standard observable-side quantity. In models of magnetar X-ray polarization, the polarization state of an emerging photon is determined not at the surface but at the polarization-limiting (adiabatic) radius $r_{\rm pl}$, where the birefringent decoupling of the two modes ceases to be adiabatic and the polarization ``freezes'' \cite{hey00,hey02,hey03,hey18}. Because $r_{\rm pl}$ typically lies far from the star, where the dipole field direction is nearly uniform across the visible surface, this mode tracking raises the predicted net polarization degree (the Heyl--Shaviv mechanism) \cite{hey02,hey03,taverna22}. Polarization-transport calculations routinely evaluate the adiabaticity criterion with the weak-field Cotton--Mouton birefringence, $\Delta n_{\rm wf}=(\alpha/30\pi)\,\xi^{2}\sin^{2}\theta$, even for surface fields exceeding $B_{\rm cr}$, where Sec.~\ref{sec:propagation} shows that this form overestimates the exact one-loop $\Delta n$ substantially. We quantify the error this induces in $r_{\rm pl}$.

We adopt the following explicit convention (Gaussian units). Along a radial ray, the local adiabaticity parameter is
\begin{equation}
\label{eq:adiabaticity}
\Gamma(r)
\equiv
\frac{\omega}{c}\,
\bigl|\Delta n(r)\bigr|\,
\ell_{B}(r),
\qquad
\ell_{B}
=
\left|\frac{d\ln B}{dr}\right|^{-1},
\end{equation}
and $r_{\rm pl}$ is defined by $\Gamma(r_{\rm pl})=1$ \cite{hey00,hey02}. Different authors place $\mathcal{O}(1)$ factors in this criterion; since $\Gamma\propto r^{-5}$ for a dipole in the weak-field regime, such factors shift $r_{\rm pl}$ only by their one-fifth power and cancel exactly in the exact-to-weak-field ratio that is our object of interest. For a centered dipole, $B(r)=B_{s}(R_{\rm NS}/r)^{3}$ with $R_{\rm NS}=10$~km, one has $\ell_{B}=r/3$, and we take $\theta=\pi/2$ (maximal birefringence; the weak-field angular dependence is $\sin^{2}\theta$). We compute $r_{\rm pl}$ twice: (a) with $\Delta n_{\rm wf}$, for which the criterion admits the closed form
\begin{equation}
\label{eq:rpl_wf}
r_{\rm pl}^{\rm wf}
=
R_{\rm NS}
\left[
\frac{\alpha}{90\pi}\,
\frac{\omega R_{\rm NS}}{c}\,
\xi_{s}^{2}
\right]^{1/5},
\qquad
\xi_{s}=\frac{B_{s}}{B_{\rm cr}},
\end{equation}
so that $r_{\rm pl}^{\rm wf}\propto B_{s}^{2/5}E^{1/5}$; and (b) with the exact one-loop finite-field $\Delta n(\xi)=n_{\parallel}-n_{\perp}$ of Eqs.~(\ref{eq:n_kappa_modes}), built from the Hurwitz-zeta/digamma expressions of Sec.~\ref{sec:theory} with $\gamma_{s}$ kept exact, evaluated with adaptive arbitrary-precision arithmetic so that the exact branch remains trustworthy deep in the weak-field tail. The weak-field series is nowhere extrapolated into $\xi>1$ in branch (b).

\begin{figure*}[t]
\centering
\includegraphics[width=0.98\textwidth]{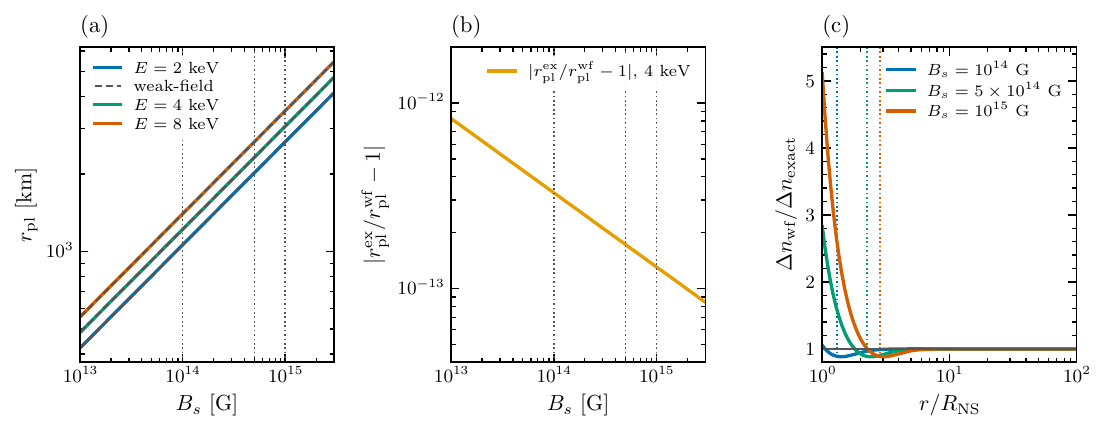}
\caption{\textbf{Finite-field correction to the polarization-limiting radius.} 
\textbf{Panel (a):} $r_{\rm pl}$ versus surface dipole field $B_{s}$ for photon 
energies $E=2$, $4$, $8$~keV, computed with the exact one-loop finite-field 
$\Delta n$ (solid) and with the weak-field Cotton--Mouton form (dashed, gray); the 
curves are indistinguishable at this scale. Vertical dotted lines mark the three 
fiducial surface fields of Table~\ref{tab:rpl}. \textbf{Panel (b):} the fractional 
difference $|r_{\rm pl}^{\rm exact}/r_{\rm pl}^{\rm wf}-1|$ at $4$~keV -- the 
internal convergence check -- which remains below $10^{-12}$ because 
$\xi(r_{\rm pl})\sim10^{-6}$ and the residual scales as $\xi^{2}(r_{\rm pl})$. 
\textbf{Panel (c):} where the weak-field formula \emph{does} fail: the local 
overestimation factor $\Delta n_{\rm wf}/\Delta n_{\rm exact}$ along the ray, for 
the three surface fields; colored dotted verticals mark $r_{\rm cr}$, the radius at 
which $B(r)=B_{\rm cr}$.}
\label{fig:rpl}
\end{figure*}

Table~\ref{tab:rpl} reports both radii for photon energies $E=2$, $4$, $8$~keV (IXPE band) and surface fields $B_{s}=10^{14}$~G (4U~0142+61-like), $5\times10^{14}$~G (1RXS~J1708$-$4009-like), and $10^{15}$~G (SGR-class); Fig.~\ref{fig:rpl} shows the corresponding curves. The central result is a null one, and it is exact in a controlled sense: the two computations of $r_{\rm pl}$ agree to a fractional difference below $10^{-12}$ for every grid point. The reason is geometric. The freeze-out radius lies at $r_{\rm pl}\simeq106$--$350\,R_{\rm NS}$, where the dipole field has decayed to $\xi(r_{\rm pl})\simeq5\times10^{-7}$--$2\times10^{-6}$; the adiabaticity criterion is local, and at such $\xi$ the exact $\Delta n$ and the Cotton--Mouton form coincide to relative order $\xi^{2}$. The convergence of the two branches for $B(r)<B_{\rm cr}$, required as an internal consistency check, is thus verified at the $10^{-12}$ level [Fig.~\ref{fig:rpl}(b)], and the residual scales as $\xi^{2}(r_{\rm pl})$ as expected.

The finite-field correction is therefore immaterial for the location of $r_{\rm pl}$ itself -- and, by extension, for the robustness of the Heyl--Shaviv enhancement of the polarization degree, which depends on $r_{\rm pl}$ lying far outside the star. Where the weak-field formula fails materially is interior to the radius $r_{\rm cr}=R_{\rm NS}\,\xi_{s}^{1/3}$ at which $B(r)=B_{\rm cr}$: $r_{\rm cr}\simeq13.1$, $22.5$, and $28.3$~km for the three surface fields considered. At the surface itself, the Cotton--Mouton form overestimates the exact $\Delta n$ by factors of $1.05$, $2.8$, and $5.1$, respectively [Fig.~\ref{fig:rpl}(c)]. Any quantity that accumulates $\Delta n$ along the ray is correspondingly affected: for the total birefringent phase $\Delta\phi=(\omega/c)\int_{R_{\rm NS}}^{\infty}\Delta n\,dr$, the weak-field formula yields $\Delta\phi_{\rm wf}/\Delta\phi_{\rm exact}=0.95$, $1.80$, and $2.86$ for $B_{s}=10^{14}$, $5\times10^{14}$, and $10^{15}$~G (the value below unity at $10^{14}$~G reflects the slight initial \emph{excess} of the exact $\Delta n$ over the $\xi^{2}$ form at $\xi\lesssim2$ visible in Fig.~\ref{fig:refractive}). We conclude that the Cotton--Mouton input is adequate at the few-percent level for $B_{s}\simeq10^{14}$~G, but overestimates near-surface birefringent transport by a factor $\sim2$ at $5\times10^{14}$~G and $\sim3$ at $10^{15}$~G; phase-sensitive observables sourced at $r\lesssim3\,R_{\rm NS}$ -- e.g.\ the QED phase delays targeted by the forecasts of Ref.~\cite{abu26} -- require the finite-field $\Delta n$ supplied here for such fields.

Several caveats bound this conclusion. The geometry is a centered dipole with radial propagation at $\theta=\pi/2$; multipolar surface fields would move $r_{\rm cr}$ but not the freeze-out region. The dielectric tensor is purely the vacuum one: no plasma contribution is included, so the vacuum-resonance mode conversion, which for magnetar-atmosphere densities occurs where plasma and vacuum contributions balance \cite{lai02,lai23,tav26}, is not modeled in this subsection (its finite-field shift is quantified in Sec.~\ref{sec:vres}); for the parameters above the resonance lies well inside $r_{\rm pl}$, so it does not affect the freeze-out analysis. Rotation of the star (which sweeps the field direction at $r_{\rm pl}$) is neglected. Finally, for $B_{s}=10^{15}$~G the surface value $\xi_{s}=22.7$ approaches the upper end of the interval $0\leq\xi\leq30$ over which the exact expressions were validated numerically in Sec.~\ref{sec:numerics_context}; there the controlling behavior is the strong-field asymptote $\Delta n\simeq(\alpha/6\pi)\,\xi$ of Sec.~\ref{sec:propagation}, so no extrapolation beyond established results is involved.

\begin{table*}[t]
\caption{Polarization-limiting radius $r_{\rm pl}$ from the adiabaticity 
criterion of Eq.~(\ref{eq:adiabaticity}), for a centered dipole with 
$R_{\rm NS}=10$~km and $\theta=\pi/2$, computed with the weak-field Cotton--Mouton 
birefringence [Eq.~(\ref{eq:rpl_wf})] and with the exact one-loop finite-field 
$\Delta n$. The last two columns give the fractional difference and the local field 
strength at freeze-out. All entries are reproducible from the companion code 
(\texttt{rpl\_magnetar.py}).}
\label{tab:rpl}
\centering
\small
\begin{tabular*}{\textwidth}{@{\extracolsep{\fill}}cccccc}
\toprule
$B_{s}$ [G] & $E$ [keV] & $r_{\rm pl}^{\rm wf}$ [km] & $r_{\rm pl}^{\rm exact}$ [km] &
$\frac{r_{\rm pl}^{\rm exact}}{r_{\rm pl}^{\rm wf}}-1$ & $\xi(r_{\rm pl})$ \\
\midrule
$10^{14}$          & 2 & $1061$ & $1061$ & $+7.5\times10^{-13}$ & $1.9\times10^{-6}$ \\
                   & 4 & $1218$ & $1218$ & $+3.3\times10^{-13}$ & $1.3\times10^{-6}$ \\
                   & 8 & $1400$ & $1400$ & $+1.4\times10^{-13}$ & $8.3\times10^{-7}$ \\
$5\times10^{14}$   & 2 & $2019$ & $2019$ & $+4.0\times10^{-13}$ & $1.4\times10^{-6}$ \\
                   & 4 & $2319$ & $2319$ & $+1.7\times10^{-13}$ & $9.1\times10^{-7}$ \\
                   & 8 & $2664$ & $2664$ & $+7.5\times10^{-14}$ & $6.0\times10^{-7}$ \\
$10^{15}$          & 2 & $2664$ & $2664$ & $+3.0\times10^{-13}$ & $1.2\times10^{-6}$ \\
                   & 4 & $3061$ & $3061$ & $+1.3\times10^{-13}$ & $7.9\times10^{-7}$ \\
                   & 8 & $3516$ & $3516$ & $+5.7\times10^{-14}$ & $5.2\times10^{-7}$ \\
\bottomrule
\end{tabular*}
\end{table*}

\subsection{Plasma--Vacuum Resonance}
\label{sec:vres}

Section~\ref{sec:rpl} established that the finite-field correction to $\Delta n$ is material only within a few stellar radii, where $B>B_{\rm cr}$. The one energy-resolved observable formed precisely in that zone, and already detected, is mode conversion at the plasma--vacuum resonance: in a magnetized atmosphere the photon modes undergo a resonance at the density where the plasma and vacuum contributions to the dielectric tensor balance \cite{lai02,lai03,vad06,lai23}. To leading order in the vacuum corrections, the resonance density and the adiabatic (mode-conversion) critical energy are \cite{lai02,lai03}
\begin{align}
\label{eq:rho_res}
\rho_{\rm res}
&\simeq
0.964\,Y_{e}^{-1}\,B_{14}^{2}\,E_{1}^{2}\,f^{-2}
\;{\rm g\,cm^{-3}},
\\
\label{eq:E_ad}
E_{\rm ad}
&=
2.52\,
\bigl[f\tan\theta_{B}\,|1-u_{i}^{-1}|\bigr]^{2/3}
\notag\\
&\quad\times
\left(\frac{5\ {\rm cm}}{H_{\rho}}\right)^{1/3}
{\rm keV},
\end{align}
with $B_{14}=B/10^{14}$~G, $E_{1}=E/1$~keV, $Y_{e}$ the electron fraction, $\theta_{B}$ the field--ray angle, $u_{i}$ the ion cyclotron parameter, $H_{\rho}$ the density scale height, and the nonadiabatic jump probability given by the Landau--Zener form $P_{\rm jump}=\exp[-(\pi/2)(E/E_{\rm ad})^{3}]$ \cite{lai03}. The vacuum input enters through $f(B)=[3\delta_{V}/(\hat q+\hat m)]^{1/2}$, where $\delta_{V}=(\alpha/45\pi)\,\xi^{2}$ and $\hat q$, $\hat m$ are the vacuum polarizability coefficients, with weak-field values $\hat q=7\delta_{V}$, $\hat m=-4\delta_{V}$, so that $f\to1$ in the Cotton--Mouton regime.

The connection to the present work is a single identity. At $\theta=\pi/2$ and to the working order $\mathcal{O}(\alpha)$ of this paper, the total vacuum anisotropy entering the resonance condition satisfies $\hat q+\hat m=2\,\Delta n+\mathcal{O}(\alpha^{2})$: this holds identically in the weak-field limit ($\hat q+\hat m=3\delta_{V}=2\Delta n_{\rm wf}$) and at finite field as well, because $\hat q$ and $\hat m$ are defined from the same $\mathcal{O}(\alpha)$ mode-index corrections whose difference is $\Delta n$. Every finite-field correction to the resonance therefore reduces to the ratio already computed exactly in this paper,
\begin{equation}
\label{eq:Rratio}
R(\xi_{\rm loc})
\equiv
\frac{\Delta n_{\rm wf}(\xi_{\rm loc})}{\Delta n_{\rm exact}(\xi_{\rm loc})},
\qquad
\xi_{\rm loc}=\frac{B_{\rm loc}}{B_{\rm cr}},
\end{equation}
evaluated at the field strength of the resonance layer, giving
\begin{align}
\label{eq:vres_shifts}
\frac{\rho_{\rm res}^{\rm exact}}{\rho_{\rm res}^{\rm wf}}
&=
\frac{1}{R},
&
f^{\rm exact}&=\sqrt{R},
\notag\\
\frac{E_{\rm ad}^{\rm exact}}{E_{\rm ad}^{\rm wf}}
&=
R^{1/3}.
&&
\end{align}
All atmosphere-dependent factors ($Y_{e}$, $H_{\rho}$, $\theta_{B}$, $u_{i}$) cancel in these ratios, which is what makes them robust statements independent of the detailed atmosphere model. We note that exact one-loop polarizability coefficients have long been available to detailed atmosphere models \cite{lai02,potekhin04}; the contribution here is the reduction of the finite-field correction to the single atmosphere-independent ratio $R$, its exact evaluation from the $\gamma_s$-normalized expressions of this work, and its source-by-source quantification against the simplified $f=1$ (Cotton--Mouton) input still common in transport calculations and analytic estimates. Because the atmosphere is only centimeters thick, $B_{\rm loc}\simeq B_{s}$ to excellent accuracy; local multipolar enhancement of the surface field would raise $\xi_{\rm loc}$ and strengthen the effect. Since $\Delta n_{\rm exact}<\Delta n_{\rm wf}$ for $\xi\gtrsim2.5$ (Fig.~\ref{fig:refractive}), the exact treatment moves the resonance to \emph{lower} density -- further out in the atmosphere -- and \emph{raises} the adiabatic energy, shifting predicted mode-conversion features toward higher photon energies relative to Cotton--Mouton-based models.

\begin{figure*}[t]
\centering
\includegraphics[width=0.85\textwidth]{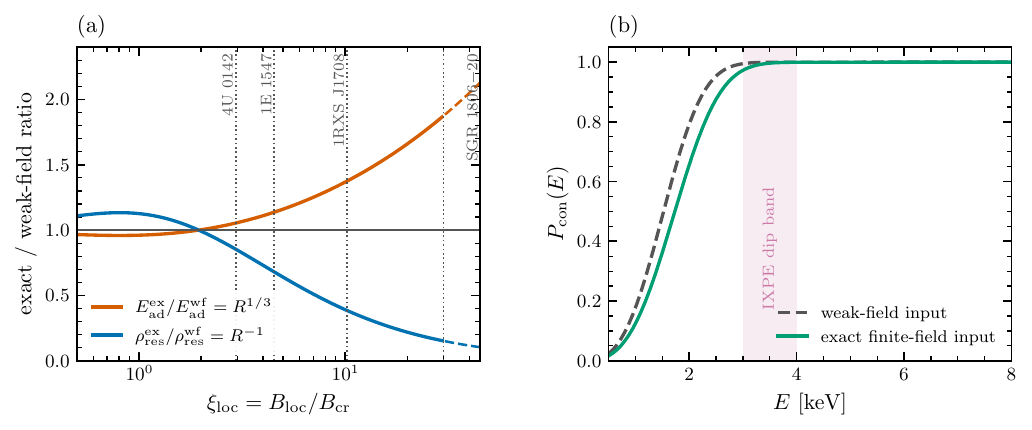}
\caption{\textbf{Finite-field correction to the vacuum resonance.} 
\textbf{Panel (a):} exact-to-weak-field shift of the resonance density 
$\rho_{\rm res}^{\rm ex}/\rho_{\rm res}^{\rm wf}=R^{-1}$ and of the adiabatic 
mode-conversion energy $E_{\rm ad}^{\rm ex}/E_{\rm ad}^{\rm wf}=R^{1/3}$ 
[Eq.~(\ref{eq:vres_shifts})] versus the local field strength 
$\xi_{\rm loc}=B_{\rm loc}/B_{\rm cr}$, with the magnetars of Table~\ref{tab:vres} 
marked; the curves are dashed beyond the validated interval $\xi\leq30$, where the 
strong-field asymptote controls. \textbf{Panel (b):} Landau--Zener mode-conversion 
probability $P_{\rm con}(E)=1-\exp[-(\pi/2)(E/E_{\rm ad})^{3}]$ for 
1E~1547.0$-$5408 ($\xi_{s}=4.53$), computed with the weak-field (dashed) and exact 
finite-field (solid) vacuum inputs for a fiducial $E_{\rm ad}^{\rm wf}=2$~keV; only 
the ratio $E_{\rm ad}^{\rm ex}/E_{\rm ad}^{\rm wf}=1.14$ is a prediction of this 
work. The shaded band marks the $3$--$4$~keV polarization-degree dip reported by the 
500~ks IXPE campaign \cite{tav26}; conversion is saturated across the band in either 
treatment, so present data cannot yet discriminate between the two inputs for this 
source.}
\label{fig:vres}
\end{figure*}

Table~\ref{tab:vres} and Fig.~\ref{fig:vres} quantify these shifts for the magnetars discussed in this work. The confrontation with 1E~1547.0$-$5408 ($\xi_{s}=4.53$) is direct: the finite-field correction raises $E_{\rm ad}$ by $13.7\%$ and lowers $\rho_{\rm res}$ by $32\%$. For any $E_{\rm ad}^{\rm wf}\lesssim 2$~keV, however, the conversion probability is already saturated ($P_{\rm con}>0.97$) across the observed $3$--$4$~keV dip band in either treatment[Fig.~\ref{fig:vres}(b)], so present IXPE data cannot discriminate between the two vacuum inputs for this source: the correction, while real, lies below current atmosphere and geometry systematics. This quantifies precisely why. The situation changes qualitatively for $\xi_{s}\gtrsim10$. For 1RXS~J1708$-$4009 -- the source identified in Ref.~\cite{abu26} as the most promising near-term vacuum-birefringence target for IXPE and eXTP -- the weak-field input misplaces the resonance density by a factor $2.6$ and the adiabatic energy by $37\%$, a shift comparable to the width of the IXPE energy bands; quantitative modeling of resonance-related spectral-polarimetric features for this source therefore \emph{requires} the finite-field $\Delta n$ supplied here. For SGR-class fields of $10^{15}$~G the corresponding factors are $5.1$ and $1.72$, and for SGR~1806$-$20 ($B_{s}\simeq2\times10^{15}$~G, $\xi_{s}=45$) they reach an order of magnitude in density and a factor $2.1$ in energy -- making the highest-field magnetars the maximal-contrast targets for a dedicated polarimetric observation. For SGR~1806$-$20 the surface field lies beyond the numerically validated interval $0\leq\xi\leq30$, and the quoted values are controlled by the strong-field asymptote $\Delta n\simeq(\alpha/6\pi)\,\xi$; they are shown dashed in Fig.~\ref{fig:vres}(a).

Two caveats bound this subsection. First, the identity $\hat q+\hat m=2\Delta n$ and Eqs.~(\ref{eq:rho_res})--(\ref{eq:E_ad}) hold to leading order in the vacuum corrections, the working order of this paper throughout. Second, the fiducial value $E_{\rm ad}^{\rm wf}=2$~keV used in Fig.~\ref{fig:vres}(b) is illustrative -- $E_{\rm ad}$ itself depends on the atmosphere through $\theta_{B}$ and $H_{\rho}$ \cite{lai03,vad06} -- and only the \emph{ratio} $E_{\rm ad}^{\rm exact}/E_{\rm ad}^{\rm wf}=R^{1/3}$ is a prediction of this work; a full conversion-probability calculation across a realistic atmosphere profile requires radiative transfer beyond the present scope. All entries are reproducible from the companion code (\texttt{vacuum\_resonance.py}).

\begin{table*}[t]
\caption{Finite-field shift of the vacuum-resonance quantities, 
Eq.~(\ref{eq:vres_shifts}), at the surface fields of the magnetars discussed in this 
work (fields from Ref.~\cite{kas14}). $R=\Delta n_{\rm wf}/\Delta n_{\rm exact}$ at 
$\xi_{\rm loc}=\xi_{s}$. The last row lies beyond the validated interval 
$\xi\leq30$ and is controlled by the strong-field asymptote (see text).}
\label{tab:vres}
\centering
\small
\begin{tabular*}{\textwidth}{@{\extracolsep{\fill}}lcccc}
\toprule
Source & $\xi_{s}$ & $R$ & $\rho_{\rm res}^{\rm ex}/\rho_{\rm res}^{\rm wf}$ &
$E_{\rm ad}^{\rm ex}/E_{\rm ad}^{\rm wf}$ \\
\midrule
4U~0142+61          & $2.95$ & $1.17$ & $0.85$ & $1.05$ \\
1E~1547.0$-$5408    & $4.53$ & $1.47$ & $0.68$ & $1.14$ \\
1RXS~J1708$-$4009   & $10.2$ & $2.60$ & $0.39$ & $1.37$ \\
SGR-class ($10^{15}$~G) & $22.7$ & $5.10$ & $0.20$ & $1.72$ \\
SGR~1806$-$20       & $45.3$ & $9.7$  & $0.10$ & $2.13$ \\
\bottomrule
\end{tabular*}
\end{table*}

%===========================================================================
\section{Conclusions}
\label{sec:conclusions}

We have investigated the effective magnetic response of photon eigenmodes in a magnetized QED vacuum within the one-loop Heisenberg--Euler framework, keeping the refractive-index normalization $\gamma_s$ exact throughout. This finite-field, mode-resolved formulation connects the photon magnetic moment, vacuum birefringence, and polarization transport in a single controlled description, valid from the weak-field Cotton--Mouton regime through $B\sim30\,B_{\rm cr}$, and avoids the factor-of-two ambiguity that arises if $\gamma_s$ is fixed before differentiation. The response is proportional to the photon energy, vanishes for propagation parallel to the field, and the photon magnetic moment is not an intrinsic property of a free photon but an effective response of the dispersion relation induced by vacuum polarization: the magnetized vacuum acts as an anisotropic optical medium, and no independent transverse magnetic-moment component is introduced.

Several consequences follow that go beyond a restatement of the underlying one-loop formulas. The normalized parallel-mode response $\widehat{\mu}_{\gamma}^{(\parallel)}(\xi)$ is strictly positive but not globally monotonic over $0\leq\xi\leq30$: it peaks at $\xi_{\rm peak}\simeq17$ ($=16.963$ within the resummed one-loop model), located exactly by the extremum condition of Eq.~(\ref{eq:A_extremum}) -- which applies to any resummed response $\mu\propto n^{-2}\,dn/d\xi$ with $n=\sqrt{1+\kappa(\xi)}$ whenever $\kappa'$ saturates while $\kappa$ grows -- and reproduced to $1.4\%$ by a closed-form estimate built from the strong-field expansion alone. The strictly truncated $\mathcal{O}(\alpha)$ response is monotonic; the maximum is generated by the resummed normalization at the $\sim2\%$ level, of the same formal order as omitted two-loop Heisenberg--Euler corrections \cite{rit75,gie17,kar19}, so its existence and approximate location, not its sub-percent profile, are the controlled statements. The finite-field correction to the magnetar polarization-limiting radius is a controlled null: $r_{\rm pl}$ computed with the exact $\Delta n$ agrees with the weak-field result to better than one part in $10^{12}$, because freeze-out occurs at $\sim10^{2}$ stellar radii where $B\ll B_{\rm cr}$ -- simultaneously an internal consistency check and a robustness statement for the Heyl--Shaviv polarization-enhancement mechanism -- while the Cotton--Mouton formula overestimates the near-surface birefringence by up to a factor $\sim5$ and the accumulated birefringent phase by up to $\sim2.9$ at $B_{s}=10^{15}$~G. The same exact $\Delta n$, propagated into the plasma--vacuum resonance, shifts the two resonance quantities through atmosphere-independent ratios. For surface fields of $2\times10^{14}$, $4.5\times10^{14}$, and $2\times10^{15}$~G, the resonance density is reduced by $32\%$, by a factor $2.6$, and by a factor $9.7$, respectively, while the adiabatic mode-conversion energy is raised by $14\%$, $37\%$, and a factor $2.13$.

The astrophysical consequences are concrete. For 1E~1547.0$-$5408, the finite-field shift lies below current systematics: the mild $3$--$4$~keV polarization-degree dip found by the 500~ks IXPE campaign \cite{tav26} is equally consistent with weak-field and exact vacuum inputs, and Sec.~\ref{sec:vres} quantifies why. For 1RXS~J1708$-$4009 -- identified as the most promising near-term target for vacuum-birefringence phase delays with IXPE and eXTP \cite{abu26} -- the weak-field input misplaces the resonance density by a factor $2.6$ and the conversion energy by $37\%$, comparable to the instrument band widths, so eXTP-era modeling of this source requires the finite-field $\Delta n$ supplied here. The highest-field magnetars offer maximal contrast: for SGR~1806$-$20 the shifts reach an order of magnitude in resonance density and a factor $2$ in conversion energy, making a dedicated polarimetric observation of an SGR-class source the sharpest available test of the finite-field regime. By contrast, the broadband polarization degree, set at the freeze-out radius, is insensitive to the correction: high observed polarization fractions test the Heyl--Shaviv mechanism, not finite-field QED. Laboratory probes remain complementary -- PVLAS ellipticity at $B\ll B_{\rm cr}$ and ATLAS light-by-light scattering in a different kinematic regime support the nonlinear-QED framework without measuring the static-field response derived here.

The main contribution of this paper is therefore a controlled finite-field, mode-resolved description of the photon magnetic response, together with quantified statements of where its corrections do and do not matter observationally. All numerical results are reproducible from the companion code accompanying this work.

\backmatter
\bmhead{Acknowledgements}
We thank Professors Frederico Della Valle, Gert Brodin, Rudolf Baier, and Victoria Kaspi for valuable suggestions. We gratefully acknowledge the late Dr. J. W. Mielniczuk for his contributions to the analytical derivations in Appendices A--C and to the numerical verification of the monotonicity analysis. S.R.V. thanks King's University College for continued support of his research work. F.A.C. thanks the Peaceful Society, Science and Innovation Foundation for support.

\section*{Statements and Declarations}

\bmhead{Funding}
No funds have been received for this study

\bmhead{Data availability}
No experimental data were generated in this study. All numerical data underlying the tables and figures can be regenerated from the companion Python scripts supplied with the submission.

\bmhead{Code availability}
The Python scripts used to evaluate the finite-field expressions, verify the numerical values, and generate all figures are supplied as supplementary source files with the submission.

\bmhead{Author contributions}
S.A. and S.R.V. developed the finite-field formulation and astrophysical applications. F.A.C. contributed to the theoretical analysis and interpretation. All authors reviewed and approved the submitted manuscript and accept responsibility for the integrity of the work.

\bmhead{Competing interests}
The authors declare that they have no competing financial or non-financial interests.

%===========================================================================
\appendix
%===========================================================================

\section{Positivity and Monotonicity}
\label{app:positivity}
\label{app:A}

For the parallel mode at $\theta=\pi/2$, the refractive index is
\begin{equation}
\label{eq:A_nperp}
n_{\parallel}
=
\left(1+\kappa_p\right)^{1/2},
\qquad
\kappa_p
=
\frac{B^2\gamma_{\mathcal{GG}}}{\gamma_s}.
\end{equation}
The corresponding photon magnetic response is
\begin{equation}
\label{eq:A_mu}
\mu_{\gamma}^{(\parallel)}
=
\frac{|\mathbf{k}|}{2(1+\kappa_p)^{3/2}}
\frac{d\kappa_p}{dB}.
\end{equation}
Thus, positivity of the parallel-mode photon magnetic response is equivalent to
\begin{equation}
\label{eq:A_condition}
\frac{d\kappa_p}{dB} \geq 0 .
\end{equation}

This form makes the role of the normalization factor $\gamma_s$ explicit. 
Unlike the simplified expression $n_\parallel\simeq1+B^2\gamma_{\mathcal{GG}}/2$, 
Eq.~(\ref{eq:A_nperp}) does not assume $\gamma_s\simeq1$.

Using the integral representation of the one-loop effective Lagrangian, the quantity 
$B^2\gamma_{\mathcal{GG}}$ may be written in terms of the dimensionless field 
$b=B/B_{\rm cr}$ as \cite{sha11}
\begin{align}
B^2\gamma_{\mathcal{GG}}
&=
\frac{\alpha}{3\pi}
\int_0^\infty
\frac{dt}{t}
\exp\left(-\frac{t}{b}\right)
{\cal I}(t),
\label{eq:A_integral}
\end{align}
where
\begin{equation}
\label{eq:A_kernel}
{\cal I}(t)
=
-\frac{3\coth t}{2t}
+
\frac{3}{2\sinh^2 t}
+
t\coth t .
\end{equation}
The kernel ${\cal I}(t)$ is non-negative for $t>0$. Therefore, the integral 
representation provides a useful way to analyze the sign of the magnetic response.

If $\gamma_s$ is held fixed at its leading zeroth-order value after the 
normalization convention has been chosen, the positivity condition reduces to the 
positivity of
\begin{equation}
\label{eq:A_derivative_fixed_gammas}
\frac{d}{dB}
\left(
B^2\gamma_{\mathcal{GG}}
\right).
\end{equation}
In the full finite-field expression, however, the derivative acts on the 
ratio $B^2\gamma_{\mathcal{GG}}/\gamma_s$,
\begin{equation}
\label{eq:A_full_derivative}
\frac{d\kappa_p}{dB}
=
\frac{d}{dB}
\left(
\frac{B^2\gamma_{\mathcal{GG}}}{\gamma_s}
\right).
\end{equation}
This is the quantity used in the numerical verification in 
Sec.~\ref{sec:experiment}.

Differentiating Eq.~(\ref{eq:A_integral}) with respect to $B$ gives
\begin{align}
\frac{d}{dB}
\left(
B^2\gamma_{\mathcal{GG}}
\right)
&=
\frac{\alpha}{3\pi B_{\rm cr}}
\int_0^\infty
dt\,
\frac{1}{b^2}
\exp\left(-\frac{t}{b}\right)
{\cal I}(t).
\label{eq:A_derivative}
\end{align}
Since ${\cal I}(t)\geq0$ for $t>0$, Eq.~(\ref{eq:A_derivative}) is non-negative. 
This establishes positivity, and in fact strict, unbounded monotonicity, of 
the un-normalized combination $B^2\gamma_{\mathcal{GG}}$ for all $B>0$: 
$d(B^2\gamma_{\mathcal{GG}})/dB>0$ everywhere, as verified numerically to hold at 
least up to $\xi\sim10^2$. For the full finite-field response 
$\widehat{\mu}_{\gamma}^{(\parallel)}\propto(1+\kappa_p)^{-3/2}\,d\kappa_p/dB$ with 
$\gamma_s$ kept explicitly, positivity is verified numerically from 
Eq.~(\ref{eq:A_full_derivative}) over the entire range $0\leq B\leq30\,B_{\rm cr}$, 
but strict monotonicity is \emph{not}: $\widehat{\mu}_{\gamma}^{(\parallel)}$ increases 
monotonically up to a broad maximum near $\xi\simeq17$ and then decreases slowly. 
The origin of this behavior is the resummation factor $(1+\kappa_p)^{-3/2}$ itself, 
not the mild field dependence of $\gamma_s$ (which deviates from unity by less than 
$0.2\%$ over the range studied): because $d\kappa_p/dB$ saturates to a finite value 
as $\xi$ grows while $\kappa_p$ itself continues to grow without bound, the 
prefactor $(1+\kappa_p)^{-3/2}$ eventually decreases faster than $d\kappa_p/dB$ 
increases, so their product develops a maximum rather than saturating monotonically 
from below. We therefore avoid claiming a global analytic proof of monotonicity for 
the normalized response; only the positivity of $d(B^2\gamma_{\mathcal{GG}})/dB$, and 
hence of $\widehat{\mu}_{\gamma}^{(\parallel)}$ itself, is established analytically here, 
while the detailed (non-monotonic) shape of $\widehat{\mu}_{\gamma}^{(\parallel)}(\xi)$ is 
established numerically (Fig.~\ref{fig:experimental}, panel b). Because the 
resummation factor retains selected $\mathcal{O}(\alpha^2)$ and higher terms while 
genuine two-loop contributions to the Heisenberg--Euler action 
\cite{rit75,gie17,kar19} are omitted, the non-monotonic shape is order-controlled 
only in the structural sense discussed in Sec.~\ref{sec:peak_analytic}: the 
existence and approximate location of the maximum are robust to corrections that 
preserve boundedness of $\kappa_p'$ and unbounded growth of $\kappa_p$, whereas its 
sub-percent profile is not.

The magnetic response is paramagnetic when
\begin{equation}
\label{eq:A_paramagnetic}
\mu_{\gamma}^{(\parallel)} > 0,
\end{equation}
or equivalently when the photon energy decreases as the external magnetic field is 
increased:
\begin{equation}
\label{eq:A_energy_condition}
\frac{d\omega_\parallel}{dB}<0.
\end{equation}
This is the physical content of the positivity condition in Eq.~(\ref{eq:A_condition}).

\subsection{Extremum Condition and Peak Estimate}
\label{app:A_peak}

We now derive the extremum condition quoted as Eq.~(\ref{eq:peak_condition_main}) in 
Sec.~\ref{sec:peak_analytic}, and the closed-form estimate of its solution. Writing 
$\widehat{\mu}_{\gamma}^{(\parallel)}(\xi)=\tfrac12(1+\kappa_p)^{-3/2}\kappa_p'(\xi)$, its 
derivative with respect to $\xi$ is
\begin{equation}
\label{eq:A_mu_derivative}
\frac{d\widehat{\mu}_{\gamma}^{(\parallel)}}{d\xi}
=
\frac12(1+\kappa_p)^{-5/2}
\left[
(1+\kappa_p)\,\kappa_p''-\frac32(\kappa_p')^2
\right].
\end{equation}
Since $(1+\kappa_p)^{-5/2}>0$ for all $\xi$, any interior critical point of 
$\widehat{\mu}_{\gamma}^{(\parallel)}(\xi)$ occurs where the bracketed factor vanishes, 
giving the extremum condition
\begin{equation}
\label{eq:A_extremum}
(1+\kappa_p)\,\kappa_p''=\frac32(\kappa_p')^2 .
\end{equation}
This condition is general: it applies to any refractive index of the resummed form 
$n=\sqrt{1+\kappa(\xi)}$ combined with the physical normalization 
$\mu\propto n^{-2}\,dn/d\xi$, independent of the specific one-loop origin of 
$\kappa(\xi)$.

A maximum is the generic outcome whenever $\kappa(\xi)$ grows without bound while 
$\kappa'(\xi)$ does not grow as fast as $\kappa(\xi)^{3/2}$. Both conditions hold 
here: Eq.~(\ref{eq:A_derivative}) shows $\kappa_p'\!\propto\! d(B^2\gamma_{\mathcal{GG}})/dB$ 
is positive for all $B$ but is bounded near its weak- and strong-field values, while 
$B^2\gamma_{\mathcal{GG}}$ itself grows without bound (numerically verified at least up 
to $\xi\sim10^2$; see also the leading strong-field term $\propto\xi$ in 
Eq.~\ref{eq:N_perp_strong}). Since $\widehat{\mu}_{\gamma}^{(\parallel)}(\xi)$ vanishes as 
$\xi\to0$ (Eq.~\ref{eq:mu_weak_modes}) and is strictly positive for all finite $\xi>0$ 
(Sec.~\ref{sec:validity}), while the prefactor 
$(1+\kappa_p)^{-3/2}$ in Eq.~(\ref{eq:A_mu_derivative}) drives it back toward smaller 
values once $\kappa_p$ becomes large, a turnover of this kind is the expected behavior 
rather than an exceptional one; Eq.~(\ref{eq:A_extremum}) simply makes this precise and 
locates it.

Solving Eq.~(\ref{eq:A_extremum}) with the exact special-function expressions for 
$\kappa_p(\xi)$, Eqs.~(\ref{eq:gammaF})--(\ref{eq:gammaGG}), gives 
$\xi_{\rm peak}=16.963$, in agreement with the maximum identified by direct numerical 
optimization of $\widehat{\mu}_{\gamma}^{(\parallel)}(\xi)$ in Sec.~\ref{sec:numerics_context}.

A closed-form estimate follows from substituting the leading-order-in-$\alpha$ 
relation $\kappa_p\simeq(\alpha/2\pi)\mathcal{N}_\parallel(\xi)$ into 
Eq.~(\ref{eq:A_extremum}), using only the strong-field expansion 
$\mathcal{N}_\parallel(\xi)$ of Eq.~(\ref{eq:N_perp_strong}) with the coefficients of 
Eq.~(\ref{eq:C_coeffs}). Differentiating this expansion twice gives, in closed form,
\begin{align}
\mathcal{N}_\parallel'(\xi)
&=
\frac23
+\frac{1+C_1(\xi)}{\xi^2}
+\frac{2C_2}{\xi^3}
+\mathcal{O}(\xi^{-4}),
\label{eq:A_Nperp_prime}
\\
\mathcal{N}_\parallel''(\xi)
&=
-\frac{3+2C_1(\xi)}{\xi^3}
-\frac{6C_2}{\xi^4}
+\mathcal{O}(\xi^{-5}),
\label{eq:A_Nperp_double_prime}
\end{align}
both verified to agree with direct numerical differentiation of 
Eq.~(\ref{eq:N_perp_strong}) to better than $10^{-9}$ relative precision over 
$10\leq\xi\leq25$. Substituting 
$\kappa_p\simeq(\alpha/2\pi)\mathcal{N}_\parallel(\xi)$, 
$\kappa_p'\simeq(\alpha/2\pi)\mathcal{N}_\parallel'(\xi)$, and 
$\kappa_p''\simeq(\alpha/2\pi)\mathcal{N}_\parallel''(\xi)$ into 
Eq.~(\ref{eq:A_extremum}) and solving numerically for its root gives
\begin{equation}
\label{eq:A_peak_closed}
\xi_{\rm peak}\simeq16.73,
\end{equation}
which agrees with the exact value $\xi_{\rm peak}=16.963$ to within $1.4\%$, using 
only quantities already derived in Sec.~\ref{sec:propagation}. This level of agreement 
is expected: the strong-field expansion of Eq.~(\ref{eq:N_perp_strong}) is most 
accurate for $\xi\gg1$, and $\xi_{\rm peak}\approx17$ lies at the edge of its regime of 
good accuracy; the residual $1.4\%$ discrepancy can be regarded as an empirical measure 
of the size of the neglected $\mathcal{O}(\xi^{-3})$ terms at this particular field 
strength.

%===========================================================================
\section{Derivative Form of the Magnetic Response}
\label{app:B}
%===========================================================================

The photon magnetic response is defined by
\begin{equation}
\label{eq:B_mu_def}
\mu_{\gamma}^{(i)}
=
-
\left(
\frac{\partial\omega_i}{\partial B}
\right)_{|\mathbf{k}|,\theta},
\qquad
i=\parallel,\perp .
\end{equation}
Using
\begin{equation}
\label{eq:B_dispersion}
\omega_i(B,\theta)
=
\frac{|\mathbf{k}|}{n_i(B,\theta)},
\end{equation}
one obtains
\begin{equation}
\label{eq:B_mu_n}
\mu_{\gamma}^{(i)}
=
\frac{|\mathbf{k}|}{n_i^2}
\left(
\frac{\partial n_i}{\partial B}
\right)_{|\mathbf{k}|,\theta}.
\end{equation}

For the parallel mode at $\theta=\pi/2$,
\begin{equation}
\label{eq:B_n_perp}
n_\parallel
=
(1+\kappa_p)^{1/2},
\end{equation}
where
\begin{equation}
\label{eq:B_kappap}
\kappa_p
=
\frac{B^2\gamma_{\mathcal{GG}}}{\gamma_s}.
\end{equation}
Differentiating Eq.~(\ref{eq:B_n_perp}) gives
\begin{equation}
\label{eq:B_dn_dB}
\frac{dn_\parallel}{dB}
=
\frac{1}{2(1+\kappa_p)^{1/2}}
\frac{d\kappa_p}{dB}.
\end{equation}
Substitution into Eq.~(\ref{eq:B_mu_n}) yields
\begin{equation}
\label{eq:B_mu_perp}
\mu_{\gamma}^{(\parallel)}
=
\frac{|\mathbf{k}|}{2(1+\kappa_p)^{3/2}}
\frac{d\kappa_p}{dB}.
\end{equation}
Equivalently,
\begin{equation}
\label{eq:B_mu_perp_gamma}
\mu_{\gamma}^{(\parallel)}
=
\frac{|\mathbf{k}|}{2(1+\kappa_p)^{3/2}}
\frac{d}{dB}
\left(
\frac{B^2\gamma_{\mathcal{GG}}}{\gamma_s}
\right).
\end{equation}

Equation~(\ref{eq:B_mu_perp_gamma}) is the finite-field expression used in 
the main text. It is preferable to a long explicit special-function formula because 
it is compact, convention-transparent, and directly connected to the refractive-index 
normalization.

The same procedure applies directly to the perpendicular mode. At $\theta=\pi/2$,
\begin{equation}
\label{eq:B_n_parallel}
n_\perp
=
\left(1-\kappa_s\right)^{-1/2},
\qquad
\kappa_s
=
\frac{B^2\gamma_{\mathcal{FF}}}{\gamma_s},
\end{equation}
so that
\begin{equation}
\label{eq:B_dn_parallel_dB}
\frac{dn_\perp}{dB}
=
\frac{1}{2}\left(1-\kappa_s\right)^{-3/2}\frac{d\kappa_s}{dB}.
\end{equation}
Substitution into Eq.~(\ref{eq:B_mu_n}), using $n_\perp^2=(1-\kappa_s)^{-1}$, gives
\begin{equation}
\label{eq:B_mu_parallel_gamma}
\mu_{\gamma}^{(\perp)}
=
\frac{|\mathbf{k}|}{2}
\left(1-\kappa_s\right)^{-1/2}
\frac{d}{dB}
\left(
\frac{B^2\gamma_{\mathcal{FF}}}{\gamma_s}
\right).
\end{equation}
Equation~(\ref{eq:B_mu_parallel_gamma}) is the finite-field companion to 
Eq.~(\ref{eq:B_mu_perp_gamma}) and is used, together with it, to evaluate 
$\widehat{\mu}_{\gamma}^{(\parallel)}-\widehat{\mu}_{\gamma}^{(\perp)}$ in 
Fig.~\ref{fig:experimental}, panel (d). As with the parallel mode, the sign and 
shape of $\mu_\gamma^{(\perp)}(B)$ are established here numerically rather than 
by a global analytic monotonicity proof.

To express the result in dimensionless form, define
\begin{equation}
\label{eq:B_xi}
\xi=\frac{B}{B_{\rm cr}},
\qquad
B_{\rm cr}=\frac{m^2}{e}.
\end{equation}
Then
\begin{equation}
\label{eq:B_d_dB}
\frac{d}{dB}
=
\frac{1}{B_{\rm cr}}
\frac{d}{d\xi}
=
\frac{e}{m^2}
\frac{d}{d\xi}.
\end{equation}
Thus the magnetic response may be written as
\begin{equation}
\label{eq:B_dimensionless_mu}
\mu_{\gamma}^{(i)}
=
\frac{e}{m}
\frac{|\mathbf{k}|}{m}
{\cal M}_i(\xi,\theta),
\end{equation}
where
\begin{equation}
\label{eq:B_dimensionless_function}
{\cal M}_i(\xi,\theta)
=
\frac{1}{n_i^2}
\left(
\frac{\partial n_i}{\partial \xi}
\right)_\theta .
\end{equation}
This form makes explicit the low-energy suppression factor 
$|\mathbf{k}|/m$ relative to the natural magnetic-moment scale $e/m$.

%===========================================================================
\section{Weak-Field Expansion and Concavity}
\label{app:C}
%===========================================================================

In the weak-field regime, the refractive indices are
\begin{subequations}
\label{eq:C_weak_indices}
\begin{align}
n_\parallel
&=
1+
\frac{\alpha}{4\pi}
\frac{14}{45}
\xi^2
\sin^2\theta
+
{\cal O}(\xi^4,\alpha^2),
\\
n_\perp
&=
1+
\frac{\alpha}{4\pi}
\frac{8}{45}
\xi^2
\sin^2\theta
+
{\cal O}(\xi^4,\alpha^2).
\end{align}
\end{subequations}
Using Eq.~(\ref{eq:B_mu_n}) and keeping only the leading one-loop terms, one obtains
\begin{subequations}
\label{eq:C_weak_mu}
\begin{align}
\mu_{\gamma}^{(\parallel)}
&=
\frac{e}{m}
\frac{|\mathbf{k}|}{m}
\frac{\alpha}{4\pi}
\frac{28}{45}
\xi
\sin^2\theta
+
{\cal O}(\xi^3,\alpha^2),
\\
\mu_{\gamma}^{(\perp)}
&=
\frac{e}{m}
\frac{|\mathbf{k}|}{m}
\frac{\alpha}{4\pi}
\frac{16}{45}
\xi
\sin^2\theta
+
{\cal O}(\xi^3,\alpha^2).
\end{align}
\end{subequations}
Therefore,
\begin{equation}
\label{eq:C_delta_mu}
\mu_{\gamma}^{(\parallel)}
-
\mu_{\gamma}^{(\perp)}
=
\frac{e}{m}
\frac{|\mathbf{k}|}{m}
\frac{\alpha}{4\pi}
\frac{4}{15}
\xi
\sin^2\theta
+
{\cal O}(\xi^3,\alpha^2).
\end{equation}

These expressions show that, in the weak-field limit, the photon magnetic 
response is positive, linear in $B$, proportional to the photon energy, and maximal 
for propagation perpendicular to the external magnetic field. The coefficient 
$28/45$ arises from differentiating the parallel-mode refractive-index correction 
proportional to $(B/B_{\rm cr})^2$; it is not a contribution from both polarization 
modes.

For a photon eigenmode with energy
\begin{equation}
\label{eq:C_energy}
\omega_i
=
\frac{|\mathbf{k}|}{n_i},
\end{equation}
the first derivative is
\begin{equation}
\label{eq:C_mu_relation}
\frac{d\omega_i}{dB}
=
-\mu_{\gamma}^{(i)} .
\end{equation}
Thus, when $\mu_{\gamma}^{(i)}>0$, the photon energy decreases with increasing 
magnetic field. The second derivative is
\begin{equation}
\label{eq:C_second_derivative}
\frac{d^2\omega_i}{dB^2}
=
-
\frac{d\mu_{\gamma}^{(i)}}{dB}.
\end{equation}
Therefore, if $d\mu_{\gamma}^{(i)}/dB>0$, the photon energy is a concave function of 
$B$:
\begin{equation}
\label{eq:C_concavity}
\frac{d^2\omega_i}{dB^2}<0.
\end{equation}

This is the correct mathematical statement: a negative second derivative 
corresponds to concavity, not convexity. In the weak-field regime, 
Eq.~(\ref{eq:C_weak_mu}) gives $d\mu_{\gamma}^{(i)}/dB>0$ at leading order, and 
therefore the photon energy is concave as a function of the external magnetic field.

\end{document}